\documentclass[english,aps, pra, twocolumn,  superscriptaddress]{revtex4-2}
\usepackage[T1]{fontenc}
\usepackage[latin9]{inputenc}
\usepackage{babel}
\usepackage{amsmath}
\usepackage{amssymb}
\usepackage{graphicx}
\usepackage{esint}
\usepackage[pdfusetitle,
 bookmarks=true,bookmarksnumbered=false,bookmarksopen=false,
 breaklinks=false,pdfborder={0 0 1},backref=false,colorlinks=false]
 {hyperref}
\hypersetup{
 colorlinks,linkcolor=blue,citecolor=blue,urlcolor=blue}

\makeatletter
\renewcommand\[{\begin{equation}} 
\renewcommand\]{\end{equation}}
\renewenvironment{align*}{\align}{\endalign}

\makeatother

\begin{document}
\global\long\def\vect#1{\overrightarrow{\mathbf{#1}}}%

\global\long\def\abs#1{\left|#1\right|}%

\global\long\def\av#1{\left\langle #1\right\rangle }%

\global\long\def\ket#1{\left|#1\right\rangle }%

\global\long\def\bra#1{\left\langle #1\right|}%

\global\long\def\tensorproduct{\otimes}%

\global\long\def\braket#1#2{\left\langle #1\mid#2\right\rangle }%

\global\long\def\omv{\overrightarrow{\Omega}}%

\global\long\def\inf{\infty}%

\title{Exact nonlinear dynamics of single band metallic systems with dissipation:
From optical to dc currents}
\author{J. M. Alendouro Pinho}
\email{up201703751@fc.up.pt}

\address{Departamento de Física e Astronomia, Faculdade de Ciências da Universidade
do Porto, Rua do Campo Alegre, s/n, 4169-007 Porto, Portugal}
\address{Centro de Física das Universidades do Minho e do Porto (CF-UM-UP)
and Laboratory of Physics for Materials and Emergent Technologies
LaPMET, University of Porto, 4169-007 Porto, Portugal}
\author{B. Amorim}
\address{Departamento de Física e Astronomia, Faculdade de Ciências da Universidade
do Porto, Rua do Campo Alegre, s/n, 4169-007 Porto, Portugal}
\address{Centro de Física das Universidades do Minho e do Porto (CF-UM-UP)
and Laboratory of Physics for Materials and Emergent Technologies
LaPMET, University of Porto, 4169-007 Porto, Portugal}
\author{Yuliy V. Bludov}
\address{Centro de Física das Universidades do Minho e do Porto (CF-UM-UP)
and Laboratory of Physics for Materials and Emergent Technologies
LaPMET, University of Porto, 4169-007 Porto, Portugal}
\author{J. M. Viana Parente Lopes}
\email{jlopes@fc.up.pt}

\address{Departamento de Física e Astronomia, Faculdade de Ciências da Universidade
do Porto, Rua do Campo Alegre, s/n, 4169-007 Porto, Portugal}
\address{Centro de Física das Universidades do Minho e do Porto (CF-UM-UP)
and Laboratory of Physics for Materials and Emergent Technologies
LaPMET, University of Porto, 4169-007 Porto, Portugal}
\begin{abstract}
The theory of open quantum systems is one of the most essential tools
for the development of quantum technologies. A particular area of
interest is in the optical response of solid state systems, where
dissipation is introduced phenomenologically through the relaxation
time approximation and its effects are usually gauged perturbatively.
Analytical exact results for driven systems under this approximation
are scarce and tipically pertain only to the stationary regime. Here,
we obtain the analytical solution for the current response of general
single-band tight-binding system driven by a uniform electric field
with generic time-dependence under the relaxation time approximation.
We explore the effects of dissipation in two limiting cases: A monochromatic
field, where we analytically deduce the effect of dissipation on High
Harmonic Generation, and a constant electric field, where a generalization
for the Esaki-Tsu equation is presented for any single-band tight-binding
system. We specify the results for a simple 2D nearest neighbours
tight-binding lattice to emphasize the effect of the scale competition
introduced by the two different neighbours in both the monochromatic
and constant field cases. Finally, we compare our exact result for
the constant field to the one obtained from the usual perturbation
theory calculation to probe the validity of the latter. 
\end{abstract}
\maketitle

\section{Introduction}

Modelling the optical response of solid state materials is a cornerstone
of modern optoelectronics. New devices are often based on desirable
optical properties that can be engineered in synthetic materials \citep{zhang_review_2021}.
This requires a fine understanding of the physics underlying the light-matter
interactions in these systems.

The investigation of optical properties in solid state systems amounts
to determining the electrical currents generated by the electromagnetic
field of an incomming light. For a long time, available light sources
were weak, such that the current response of most materials could
be described within the framework of linear response. With the advent
of lasers \citep{strickland_compression_1985}, however, hitherto
ignored nonlinear currents became a central feature of a new class
of physical phenomena. Since then, strong-field phenomena in quantum
systems became an active field of research \citep{aoki_nonequilibrium_2014,oka_floquet_2019,eckardt_colloquium_2017,basov_towards_2017,de_la_torre_colloquium_2021,morimoto_geometric_2023,silva_high_2019}.
A hallmark of nonlinear response is that in such phenomena, the system
responds at frequencies that are multiples of the fundamental driving
frequency. Second and third harmonic generation \citep{mund_high-resolution_2018,kumar_third_2013},
Monochromatic wave mixing and high-harmonic generation\citep{silva_high_2019,ghimire_high-harmonic_2019}
are examples of such nonlinear effects.

When the driving field becomes strong, environmental dissipation effects
play a vital role in determining the out-of-equilibrium state of the
system.This problem has a long history of research, which has lead
to the development of several frameworks to include dissipative mechanisms
in quantum theories \citep{weiss_quantum_2012,barreiro_transport_2009,vandecasteele_current-voltage_2010,fang_high-field_2011,li_nonequilibrium_2018,okamoto_nonequilibrium_2007,sugimoto_field-induced_2008,heidrich-meisner_nonequilibrium_2010,ikeda_nonequilibrium_2021}.
One widely used theoretical tool to formulate the dynamics of open
systems is the Keldysh formalism \citep{sadovskii_diagram_2023},
which utilises the nonequilibrium Green\textquoteright s function
(NEGF) \citep{aoki_nonequilibrium_2014,blandin_localized_1976,jauho_time-dependent_1994,onoda_theory_2006,han_nonequilibrium_2018,tsuji_correlated_2008}
as a fundamental quantity rather than the wave function. The NEGF
formalism is ideally suited for the development of perturbative and
self-consistent approximations. An alternative class of methods to
describe open-quantum systems are density matrix equations of motions,
typically referred to as Quantum Master Equations (QMEs)\citep{breuer_theory_2002,alicki_quantum_2007}.
Such equations can sometimes be reduced to the so called Lindblad
equation, which describe Markovian evolution \citep{breuer_theory_2002}.
The simplest, but still useful, description of dissipation is provided
by the Relaxation Time Approximation (RTA), which assumes that in
the absence of drive the system relaxes to a known equilibrium state,
in a time-scale determined by a relaxation rate\citep{sato_nonlinear_2021,sato_light-induced_2019,nuske_floquet_2020,boyd_nonlinear_2008,yamada_density_2014}.
The RTA has been widely used in conjunction with the Boltzmann equation
\citep{peres_optical_2014,mikhailov_non-linear_2007} and the semiconductor
Bloch equations, on top of which perturbation theory schemes have
been developed\citep{passos_nonlinear_2018,passos_nonlinear_2021,ventura_gauge_2017,sipe_nonlinear_1993,sipe_second-order_2000}.
Despite its phenomenological nature, which can bring about unphysical
behaviour in multiband systems \citep{terada_limitations_2024}, the
RTA is suficient to capture a broad range of rich phenomena. Examples
of these are the generation of direct 

An example of nonlinear optical effect is the production of direct
currents driven by optical excitation, a class of responses that is
collectively known as photogalvanic effects (PGEs) \citep{ventura_second_2020,passos_nonlinear_2021,sodemann_quantum_2015,chan_photocurrents_2017,de_juan_quantized_2017,cheng_intraband_2019}.
This class of responses can be predicted with perturbation theory
(on the electromagnetic field) within RTA \citep{sipe_second-order_2000},
but the direct currents show anomalous dependencies on the relaxation
time \citep{fregoso_jerk_2018,ventura_second_2020,cheng_intraband_2019}.
In particular, they diverge in the collisionless limit, which questions
the validity of standard perturbation theory. Nonperturbative calculations
with Floquet-Keldysh formalism instead predict that these currents
converge to a finite value in this limit \citep{matsyshyn_rabi_2021}.
Other than this, direct comparisons between standard perturbative
and nonperturbative methods, probing the validty of perturbation theory,
are lacking. 

The existence of exact analytical results, even when obtained for
some simplified model, is invaluable in physics, for their rarity
and since they enable comparison with both experimental results and
theoretical methods obtained via approximate and/or numerical methods.
This is also true in the study of solid-state optoelectronic systems,
due to the challenges present by treating both strong fields and dissipation
in an adequate way. Although analytical solutions for certain driven
dissipative systems have been derived, they typically apply only to
the stationary regime under a constant external field \citep{han_solution_2013,kolovskyAtomicCurrentOptical2008,minot_quantum_2004}.
In this paper, we present the exact solution for the time-resolved
current response of a electronic system described by a single-band
tight-binding system, subject to a uniform electric field with arbitrary
time-dependency, described within a QME under the RTA. The paper is
organized as follows: in Section \ref{sec:Solution} we detail the
theoretical framework and model used. We outline the steps taken to
obtain the solution for the current and deduce the analytical exact
expressions for the current's time evolution in two limiting cases:
a monochromatic field and a constant field, both turned-on at an initial
time $t=0$. In Section \ref{sec:Results} we obtain the analytical
expression for the strong-field response of the system, discussing
both: (i) the effect of dissipation on High Harmonic Generation (HHG),
and discuss the emergence of even harmonics in centrosymmetric systems;
and (ii) the steady-state DC current obtained due to a constant electric
field, where a generalization for the Esaki-Tsu equation is presented,
valid for any single-band tight-binding model and arbitrary spatial
dimensionality. In Section \ref{sec:Comparison-with-Perturbation},
we compare the perturbative results with the exact ones. Finally,
we state our conclusions in Section \ref{sec:Conclusion}.

\section{Model and analytical solution\label{sec:Solution}}

One common approach to study the current response of condensed matter
systems with dissipation is via a von-Neumann like equation for the
1-particle reduced density matrix (1-RDM) for electrons, $\rho(t)$,
which reads
\begin{equation}
\frac{\partial\rho\left(t\right)}{\partial t}=-\frac{i}{\hbar}\left[H\left(t\right),\rho\left(t\right)\right]-\left.\frac{\partial\rho\left(t\right)}{\partial t}\right|_{\text{scat}},\label{eq:QDM_eq_scatt}
\end{equation}
where $H(t)$ is the time-dependent Hamiltonian for independent electrons,
and $\left.\frac{\partial\rho\left(t\right)}{\partial t}\right|_{\text{scat}}$
includes the scattering induced by electron-electron, electron-phonon
and electron-impurity interactions \citep{cheng_third-order_2015}.
This type of equation is also referred to as a semiconductor Bloch
equation. In principle, this type of equationcan be obtained from
well-established treatments of many-particle systems, such as the
many-particle density matrix framework \citep{haug_quantum_2009,malic_microscopic_2011}
or the Keldysh Green function method \citep{haug_quantum_2008}. The
simplest approximation to the scattering term is the so-called Relaxation
Time Approximation (RTA), in which Eq.~(\ref{eq:QDM_eq_scatt}) reduces
to\citep{sato_nonlinear_2021,sato_light-induced_2019,nuske_floquet_2020,boyd_nonlinear_2008,yamada_density_2014}

\begin{equation}
\frac{d\rho(t)}{dt}=-\frac{i}{\hbar}\left[H\left(t\right),\rho\left(t\right)\right]-\gamma\left(\rho(t)-\rho^{\text{eq}}(t)\right),\label{eq:QDM_many_scatts}
\end{equation}
where $\gamma$ is a relaxation rate and $\rho^{\text{eq}}(t)$ is
a (possibly time-dependent) equilibrium 1-RDM, towards which the system
will relax in the absence of external drive.

We will focus of crystalline electronic systems, whose electronic
dispersion relation can be described by a single isolated band. We
describe the coupling of the electrons to a uniform electric field,
via Peierls substitution, $\mathbf{k}\rightarrow\mathbf{k}+e\mathbf{A}\left(t\right)/\hbar$,
where $\mathbf{A}\left(t\right)=-\int_{0}^{t}\mathbf{E}\left(t'\right)dt'$
is the vector potential, such that $\mathbf{E}(t)=-\partial_{t}\mathbf{A}(t)$.
To simplify notation, we will define $\mathbf{k}\left(t\right)\equiv\mathbf{k}+e\mathbf{A}\left(t\right)/\hbar$.
As such, the time-dependent Hamiltonian is written in the Bloch basis
as
\[
H(t)=\sum_{\mathbf{k}}\varepsilon\left(\mathbf{k}+e\mathbf{A}\left(t\right)/\hbar\right)\ket{\mathbf{k}}\bra{\mathbf{k}},
\]
where $\mathbf{k}$ is the electron's Bloch momentum, $\ket{\mathbf{k}}$
is a Bloch state and $\varepsilon\left(\mathbf{k}\right)$ is the
single band dispersion relation. We can write $\varepsilon\left(\mathbf{k}\right)$
as a Fourier series,
\[
\varepsilon\left(\mathbf{k}\right)=\sum_{\mathbf{l}}t_{\mathbf{l}}e^{i\mathbf{k}\cdot\mathbf{l}},
\]
where $t_{\mathbf{l}}$ are real-space hoppings and $\mathbf{l}$
is the set of coordinates of the l-th neighbour. We will assume that
the system has time-reversal symmetry, such that the $t_{\mathbf{l}}$
are real, and $t_{\mathbf{l}}=t_{-\mathbf{l}}^{*}$ to preserve the
hermiticity of the Hamiltonian. The total current operator is obtained
as $\mathbf{J}\left(t\right)=\left.\delta H/\delta\mathbf{A}\right|_{\mathbf{A}=\mathbf{A}\left(t\right)}$.
In the Bloch basis we can write $\mathbf{J}\left(t\right)=\sum_{\mathbf{k}}\mathbf{J}_{\mathbf{k}}\left(t\right)\ket{\mathbf{k}}\bra{\mathbf{k}}$,
where
\begin{equation}
\mathbf{J}_{\mathbf{k}}\left(t\right)=\frac{e}{\hbar}\sum_{\mathbf{l}}it_{\mathbf{l}}\mathbf{l}e^{i\left(\mathbf{k}+e\mathbf{A}\left(t\right)/\hbar\right)\cdot\mathbf{l}}.\label{eq:curren_operator-1}
\end{equation}
We assume the system is initially in a thermalized state described
by the Fermi-Dirac distribution. The density matrix of the initial
distribution is a matrix diagonal in the Bloch momentum basis, given
by
\[
\rho_{\mathbf{k}}^{\text{eq}}=\frac{1}{e^{\beta\left(\varepsilon\left(\mathbf{k}\right)-\mu\right)}+1}.
\]
where $\beta$ is the inverse temperature and $\mu$ is the chemical
potential. In the long time limit, we can have a zero external electric
field, but a non-zero vector potential. The equilibrium density matrix,
$\rho^{\text{eq}}(t)$, appearing in the relaxation term of Eq.~(\ref{eq:QDM_many_scatts})
should take this into account by setting $\rho_{\mathbf{k}}^{\text{eq}}(t)=\rho_{\mathbf{k}+e\mathbf{A}\left(t\right)/\hbar}^{\text{eq}}$
, where we shift the Bloch momentum of the unperturbed equilibrium
distribution by $e\mathbf{A}\left(t\right)$ \citep{passos_nonlinear_2018}.
This shift guarantees that $\lim_{t\rightarrow\infty}\rho_{\mathbf{k}}(t)=\rho_{\mathbf{k}+e\mathbf{A}\left(t\right)/\hbar}^{\text{eq}}$
is solution of Eq.~(\ref{eq:QDM_many_scatts}) provided $\lim_{t\rightarrow\infty}\mathbf{E}(t)=\mathbf{0}$.

We will now explain how to obtain the exact solution for $\rho_{\mathbf{k}}(t)$.

\subsection{General solution}

In orther to solve Eq.~(\ref{eq:QDM_many_scatts}), we start by making
the crucial observation that for a spatially uniform electric field
and for an initial density matrix diagonal in the Bloch momentum basis,
$\rho(t)$ will remain diagonal in the Bloch basis, $\rho\left(t\right)=\sum_{\mathbf{k}}\rho_{\mathbf{k}}\left(t\right)\ket{\mathbf{k}}\bra{\mathbf{k}}$.
Therefore, for a single-band system, Eq.~(\ref{eq:QDM_many_scatts})
simplifies to
\begin{equation}
\frac{d\rho_{\mathbf{k}}(t)}{dt}=-\gamma\left(\rho_{\mathbf{k}}(t)-\rho_{\mathbf{k}}^{\text{eq}}(t)\right).\label{eq:DM_equation_simple}
\end{equation}
We will focus on the expectation value of the total current,$\left\langle \mathbf{J}\right\rangle \left(t\right)=\sum_{\mathbf{k}}\rho_{\mathbf{k}}\left(t\right)\mathbf{J}_{\mathbf{k}}\left(t\right)$.
From Eq. (\ref{eq:curren_operator-1}), one can write $\left\langle \mathbf{J}\right\rangle \left(t\right)$
current as
\begin{equation}
\left\langle \mathbf{J}\right\rangle \left(t\right)=i\frac{e}{\hbar}\sum_{\mathbf{l}}t_{\mathbf{l}}\mathbf{l}\tilde{\mathcal{J}}_{\mathbf{l}}\left(t\right),\label{eq:av_current}
\end{equation}
where $\tilde{\mathcal{J}}_{\mathbf{l}}\left(t\right)\equiv\sum_{\mathbf{k}}\rho_{\mathbf{k}}\left(t\right)e^{i\mathbf{k}\left(t\right)\cdot\mathbf{l}}$.
To obtain or analytical solution for $\left\langle \mathbf{J}\right\rangle \left(t\right)$,
we will use Eq.~(\ref{eq:DM_equation_simple}) to obtain a differential
equation for for $\tilde{\mathcal{J}}_{\mathbf{l}}\left(t\right)$,
which we will then solve. Multiplying Eq. \ref{eq:DM_equation_simple}
by $e^{i\mathbf{k}\left(t\right)\cdot\mathbf{l}}$ and performing
the sum over $\mathbf{k}$ allows us to write
\[
\frac{d\tilde{\mathcal{J}}_{\mathbf{l}}\left(t\right)}{dt}-\frac{e}{\hbar}\left(\frac{d\mathbf{A}\left(t\right)}{dt}\cdot\mathbf{l}\right)\tilde{\mathcal{J}}_{\mathbf{l}}\left(t\right)=-\gamma\left(\tilde{\mathcal{J}}_{\mathbf{l}}\left(t\right)-C_{\mathbf{l}}\right),
\]
where we used the fact that $de^{i\mathbf{k}\left(t\right)\cdot\mathbf{l}}/dt=\frac{d\mathbf{k}\left(t\right)}{dt}\cdot\mathbf{l}e^{i\mathbf{k}\left(t\right)\cdot\mathbf{l}}$
and that $\frac{d\mathbf{k}\left(t\right)}{dt}=\frac{e}{\hbar}\frac{d\mathbf{A}\left(t\right)}{dt}$
in the LHS through partial differentiation. Note that the equation
contains the quantity $C_{\mathbf{l}}\equiv\frac{V_{d}}{\left(2\pi\right)^{d}}\int_{FBZ}d\mathbf{k}\rho_{\mathbf{k}\left(t\right)}^{\text{eq}}e^{i\mathbf{k}\left(t\right)\cdot\mathbf{l}}=\frac{V_{d}}{\left(2\pi\right)^{d}}\int_{FBZ}d\mathbf{k}\rho_{\mathbf{k}}^{\text{eq}}e^{i\mathbf{k}\cdot\mathbf{l}}$
which is constant in time, where $V_{d}$ is the volume of the unit
cell and $d$ is the dimension of the system. The fact that the quantities
$\tilde{\mathcal{J}}_{\mathbf{l}}\left(t\right)$ are proportional
to a sum of exponentials that contain all time dependency allows a
set of decoupled, ordinary and linear differential equations for $\tilde{\mathcal{J}}_{\mathbf{l}}\left(t\right)$.
The solution to these equations is:
\[
\tilde{\mathcal{J}}_{\mathbf{l}}\left(t\right)=C_{\mathbf{l}}e^{i\frac{e}{\hbar}\mathbf{A}\left(t\right)\cdot\mathbf{l}-\gamma t}\left(1+\int_{0}^{t}e^{\left(-i\frac{e}{\hbar}\mathbf{A}\left(t'\right)\cdot\mathbf{l}+\gamma t'\right)}dt'\right).
\]
We can now construct the solution for the average current by simply
summing over these terms $\tilde{\mathcal{J}}_{\mathbf{l}}\left(t\right)$
as stated in Eq. \ref{eq:av_current}. Because we have fixed $t_{\mathbf{l}}=t_{-\mathbf{l}}$,
we can organize the sum in terms of the quantity $2iC_{\mathbf{l}}\mathcal{J}_{\mathbf{l}}\left(t\right)\equiv\left(\tilde{\mathcal{J}}_{\mathbf{l}}\left(t\right)-\tilde{\mathcal{J}}_{-\mathbf{l}}\left(t\right)\right)$.
The full solution is therefore 
\[
\left\langle \mathbf{J}\right\rangle \left(t\right)=-\frac{2e}{\hbar}\sum_{\mathbf{l}>0}t_{\mathbf{l}}\mathbf{l}C_{\mathbf{l}}\mathcal{J}_{\mathbf{l}}\left(t\right),
\]
where $\sum_{\mathbf{l}>0}$ is notation for a sum over only one of
the pair of neighbours at positions $\mathbf{l}$ and $-\mathbf{l}$
and
\begin{equation}
\begin{aligned}\mathcal{J}_{\mathbf{l}}\left(t\right) & =e^{-\gamma t}\left\{ \sin\left(\frac{e}{\hbar}\mathbf{A}\left(t\right)\cdot\mathbf{l}\right)\right.\\
 & \left.+\gamma\int_{0}^{t}dt'e^{\gamma t'}\sin\left[\frac{e}{\hbar}\left(\mathbf{A}\left(t\right)-\mathbf{A}\left(t'\right)\right)\cdot\mathbf{l}\right]\right\} .
\end{aligned}
\label{eq:exact_sol_general}
\end{equation}
Is the contribution of each pair of neighbours at positions $\mathbf{l}$
and $-\mathbf{l}$.

\subsection{Monochromatic field}

A common interest in the study of the optical response of materials
is to study the high harmonic generation (HHG) produced as a response
to a monochromatic field. By choosing $\mathbf{E}\left(t\right)=-\mathbf{E}\Theta\left(t\right)\cos\left(\Omega t+\phi\right)$
and therefore $\mathbf{A}(t)=\frac{\mathbf{E}}{\Omega}\Theta\left(t\right)\left(\sin\left(\Omega t+\phi\right)-\sin\phi\right)$,
it is possible to calculate the intensity of the generated harmonics
of the quasi-steady state regime of a system evolving according to
Eq. (\ref{eq:QDM_many_scatts}) using Eq. \ref{eq:exact_sol_general}.
Using the Jacobi-Anger expansion $e^{iz\sin\theta}\equiv\sum_{n=-\infty}^{\infty}J_{n}(z)e^{in\theta}$,
where $J_{n}\left(z\right)$ is the Bessel function of the first kind
of order $n$, it is a straightforward calculation to show that from
Eq. \ref{eq:exact_sol_general} one obtains
\begin{equation}
\begin{alignedat}{1} & \mathcal{J}_{\mathbf{l}}\left(t\right)=-\sum_{j=0}^{\infty}\left|\Lambda_{2j+1}^{\mathbf{l}}\right|\sin\left(\left(2j+1\right)\left(\Omega t+\phi\right)+\psi_{2j+1}^{\mathbf{l}}\right)\\
 & +e^{-\gamma t}\left\{ \chi_{2j+1}^{\mathbf{l},1}\sin\left(\left(2j+1\right)\left(\Omega t+\phi\right)\right)+\chi_{2j}^{\mathbf{l},2}\cos\left(2j\left(\Omega t+\phi\right)\right)\right\} ,
\end{alignedat}
\label{eq:exact_sol_mono}
\end{equation}
With $\psi_{2j+1}^{\mathbf{l}}=\arg(\Lambda_{2j+1}^{\mathbf{l}})$
and
\begin{equation}
\begin{aligned} & \begin{aligned}\Lambda_{j}^{\mathbf{l}}\end{aligned}
=2\sum_{m=0}^{\infty}\frac{J_{m}^{\mathbf{l}}}{1+\delta_{0m}}\left(\frac{J_{m+j}^{\mathbf{l}}}{1+im\frac{\Omega}{\gamma}}-\frac{J_{m-j}^{\mathbf{l}}}{1-im\frac{\Omega}{\gamma}}\right),\\
 & \chi_{j}^{\mathbf{l},1}=2\cos\left(\frac{\omega_{B}^{\mathbf{l}}}{\Omega}\sin\left(\phi\right)\right)J_{j}^{\mathbf{l}}\\
 & \quad\quad\quad\quad\quad-4\sum_{m\,\text{even}}^{\infty}\frac{J_{j}^{\mathbf{l}}J_{m}^{\mathbf{l}}\left(\cos m\phi+m\frac{\Omega}{\gamma}\sin m\phi\right)}{\left(1+\delta_{0m}\right)\left(1+\left(m\frac{\Omega}{\gamma}\right)^{2}\right)},\\
 & \chi_{j}^{\mathbf{l},2}=2\sin\left(\frac{\omega_{B}^{\mathbf{l}}}{\Omega}\sin\left(\phi\right)\right)J_{j}^{\mathbf{l}}\\
 & \quad\quad\quad-\frac{4}{1+\delta_{0j}}\sum_{m\,\text{odd}}^{\infty}\frac{J_{j}^{\mathbf{l}}J_{m}^{\mathbf{l}}\left(m\frac{\Omega}{\gamma}\cos\left(m\phi\right)-\sin\left(m\phi\right)\right)}{1+\left(m\frac{\Omega}{\gamma}\right)^{2}},
\end{aligned}
\label{eq:coefs_mono}
\end{equation}
Where we have defined $J_{n}^{\mathbf{l}}\equiv J_{n}\left(\left|\omega_{B}^{\mathbf{l}}/\Omega\right|\right)$
to simplify notation, with $\omega_{B}^{\mathbf{l}}=\frac{e}{\hbar}El\cos\phi_{\mathbf{l}}$,
$l\equiv\left|\mathbf{l}\right|$, $\cos\phi_{\mathbf{l}}\equiv\mathbf{l}\cdot\mathbf{A}\left(t\right)/\left(lA\left(t\right)\right)$
and $A\left(t\right)\equiv\left|\mathbf{A}\left(t\right)\right|$.
Each contribution to the current $\mathcal{J}_{\mathbf{l}}\left(t\right)$
is composed of a persistent oscillating term and a transient term,
which depends on the sudden switching of the electric field, that
decays exponentially in time as $e^{-\gamma t}$. It is worth noting
at this point that the solution displays inversion symmetry, since
inverting the electric field also results in the complete inversion
of the current (this is made evident by simply performing the change
$\phi\rightarrow\phi+\pi$). This makes sense, since a single-band
system inherently displays inversion symmetry. The Taylor expansion
of the Bessel functions in the electric field also allows us to gauge
this. Because they have a very particular form of expansion $J_{\alpha}\left(z\right)=\sum_{n=0}^{\infty}C_{n}z^{2n+\alpha}$,
it is not hard to see from the coefficients in Eq.~\ref{eq:coefs_mono}
that only odd powers of the electric field amplitude appear, which
means that the current is inverted if the electric field is inverted.

The persistent current is comprised of a sum of odd harmonics of the
electric field fundamental frequency $\Omega$, with corresponding
amplitude $\begin{aligned}\Lambda_{j}^{\mathbf{l}}\end{aligned}
$. Curiously, the transient contribution presents even harmonics of
the electric field frequency, with non-vanishing amplitudes $\chi_{j}^{\mathbf{l},2}$.
In fact, even in the zero dissipation limit $\gamma\rightarrow0$
we can get persistent even harmonic oscillations for $\phi\neq0$.
In this limit, the coefficients in Eq. \ref{eq:coefs_mono} are
\begin{equation}
\begin{aligned} & \begin{aligned}\Lambda_{j}^{\gamma=0,\mathbf{l}}\end{aligned}
=0,\\
 & \chi_{j}^{\gamma=0,\mathbf{l},1}=2\cos\left(\frac{\omega_{B}^{\mathbf{l}}}{\Omega}\sin\left(\phi\right)\right)J_{j}^{\mathbf{l}},\\
 & \chi_{j}^{\gamma=0,\mathbf{l},2}=2\sin\left(\frac{\omega_{B}^{\mathbf{l}}}{\Omega}\sin\left(\phi\right)\right)J_{j}^{\mathbf{l}}.
\end{aligned}
\label{eq:coefs_nodiss}
\end{equation}
Notice that now, in the clean limit $\gamma\rightarrow0$, the second
and third terms of Eq. \ref{eq:exact_sol_mono} do not decay and actually
dominate the response since the first term vanishes in this limit.
It is evident from Eq. \ref{eq:coefs_nodiss} that one can generate
even harmonics for $\phi\neq0$. Although the emergence of even harmonics
is usually thought to be prohibited in centrosymmetric systems, this
effect is solely due to the sudden switching of the electric field.
If the switching is rather done adiabatically, the even harmonic contributions
vanish (see Appendix \ref{subsec:Even-harmonic-Vanishing}). By introducing
dissipation, the system eventually loses memory of the electric field's
initial condition and the even harmonics vanish. Furthermore, for
$\phi=0$, the response is constituted solely of odd harmonics of
amplitude $\chi_{2j+1}^{\gamma=0,\mathbf{l},1}=2J_{2j+1}^{\mathbf{l}}$
, as reported in Ref. \citep{ghimire_observation_2011}.

The introduction of dissipation not only changes the amplitudes of
the harmonics as according to Eq. \ref{eq:coefs_mono}, it also introduces
phase differences between the harmonics. The total current at long-times
$\left\langle \mathbf{J}\right\rangle ^{\text{pers}}\left(t\right)$,
which we define as the oscillatory current originating when all transient
contributions vanish, is obtained by summing all the neighbors contributions
to the quasi-stationary regime, giving us
\[
\left\langle \mathbf{J}\right\rangle ^{\text{pers}}\left(t\right)=2\frac{e}{\hbar}\sum_{j=0}^{\infty}\mathbf{\Lambda}_{2j+1}^{T}\sin\left(\left(2j+1\right)\Omega t+\psi_{2j+1}\right),
\]
where $\mathbf{\Lambda}_{2j+1}^{T}=\left|\sum_{\mathbf{l}>0}C_{\mathbf{l}}t_{\mathbf{l}}\mathbf{l}\Lambda_{2j+1}^{\mathbf{l}}\right|$
and $\psi_{2j+1}=\arg(\sum_{\mathbf{l}>0}C_{\mathbf{l}}t_{\mathbf{l}}\mathbf{l}\Lambda_{2j+1}^{\mathbf{l}})$.
The effect of the phase differences across the different hopping contributions
is taken into account in $\mathbf{\Lambda}_{2j+1}^{T}$.

\subsection{Constant field}

For a constant field $\mathbf{E}=\mathbf{E}\Theta\left(t\right)$
turned on at $t=0$, Eq.~(\ref{eq:exact_sol_general}) becomes
\begin{equation}
\begin{aligned}\mathcal{J}_{\mathbf{l}}\left(t\right)=\frac{\omega_{B}^{\mathbf{l}}/\gamma}{1+\left(\omega_{B}^{\mathbf{l}}/\gamma\right)^{2}}\left\{ 1-e^{-\gamma t}\left(\frac{\omega_{B}^{\mathbf{l}}}{\gamma}\sin\left(\omega_{B}^{\mathbf{l}}t\right)+\cos\left(\omega_{B}^{\mathbf{l}}t\right)\right)\right\} .\end{aligned}
\label{eq:const_field_sol}
\end{equation}
We have a transient oscillating contribution at frequency $\omega_{B}^{\mathbf{l}}$
that decays exponentially and a constant value given by $\left(\omega_{B}^{\mathbf{l}}/\gamma\right)/\left(1+\left(\omega_{B}^{\mathbf{l}}/\gamma\right)^{2}\right)$.
Summing the contributions from all hopping terms to the stationary
current $\mathbf{J}^{\text{s}}\equiv\lim_{t\rightarrow\infty}\left\langle \mathbf{J}\right\rangle \left(t\right)$,
we obtain
\begin{equation}
\mathbf{J}^{\text{s}}=-2\frac{e}{\hbar}\sum_{\mathbf{l}>0}\mathbf{l}t_{\mathbf{l}}C_{\mathbf{l}}\frac{\omega_{B}^{\mathbf{l}}/\gamma}{1+\left(\omega_{B}^{\mathbf{l}}/\gamma\right)^{2}},\label{eq:general_esaki_tsu}
\end{equation}
which as we will see is a generalization of the Esaki-Tsu relation
\citep{esaki_superlattice_1970}. The physical meaning of this is
easier to unveil when considering a one dimensional tight binding
chain with nearest neighbour hoppings. In that case we only have contribution
from neighbours located at $l=1$ and the frequency $\omega_{B}^{\mathbf{l}}=\frac{e}{\hbar}E$
reduces to the Bloch frequency. The transient time is therefore composed
of decaying Bloch Oscillations (BOs). These decaying BOs have also
been predicted to occur in two-terminal mesoscopic setups subject
to strong electric fields \citep{pinho_bloch_2023}, where the BOs
decay is caused by the coupling of the mesoscopic system to external
leads. If no dissipation was present, $\gamma=0$, then Eq. \ref{eq:const_field_sol}
would give a total current of $J\left(t\right)=-2t_{1}C\sin\left(\omega_{B}t\right)$,
which are the expected persistent BOs predicted for the one-dimensional
tight binding chain \citep{Bloch29,Zener1934}. The constant, long-time
current for the chain is $J^{\text{s}}=-2Ct_{1}\frac{E/\gamma}{\left(\frac{\hbar}{e}\right)^{2}+\left(E/\gamma\right)^{2}},$
which is the Esaki-Tsu relation that has been already obtained for
one-dimensional nearest-neighbour tight binding lattices \citep{kolovskyAtomicCurrentOptical2008,han_solution_2013,minot_quantum_2004}.
The result in Eq. \ref{eq:general_esaki_tsu} is the generalization
of the Esaki-Tsu formula for an arbitrary single-band tight-binding
system.

\section{Results\label{sec:Results}}

\begin{figure}
\centering{}\includegraphics[width=1\columnwidth]{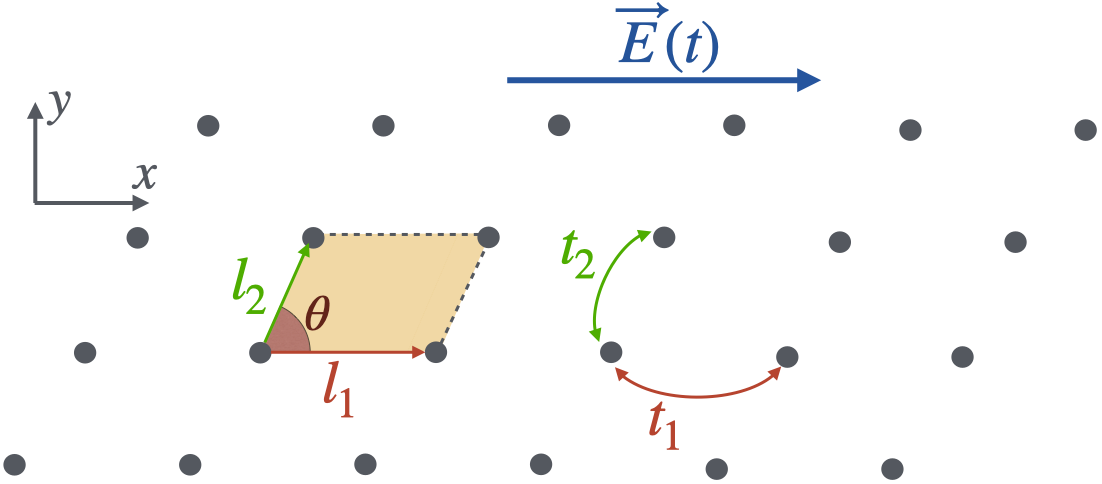}\caption{\label{fig:2D_model} Two-Dimensional tight binding lattice used.
Each site has two neighbors with hoppings $t_{1}$ and $t_{2}$ placed
at distances $l_{1}$ and $l_{2}$ from the said site and at an angle
$\theta$ with each other. The electric field lies along the direction
of the first neighbor, which will be considered the x direction. The
shaded region denotes the unit cell.}
\end{figure}

In this section, we showcase the analytical results for the monochromatic
and constant electric field cases by analysing the total current $\left\langle \mathbf{J}\right\rangle \left(t\right)$
of a two-dimensional tight binding lattice where each site has two
neighbors with hoppings $t_{1}$ and $t_{2}$ placed at distances
$l_{1}$ and $l_{2}$ from said site and at an angle $\theta$ with
each other, represented in Fig. \ref{fig:2D_model} \ref{fig:2D_model}.
 The electric field is considered to lay along the direction of the
first neighbor which lies in the x axis and we also measure the current
along this direction. The presence of an additional hopping gives
rise to competing scales that generate interesting phenomena \citep{ghimire_observation_2011}.

\subsection{Monochromatic field}

\begin{figure}
\centering{}\includegraphics[width=1\columnwidth]{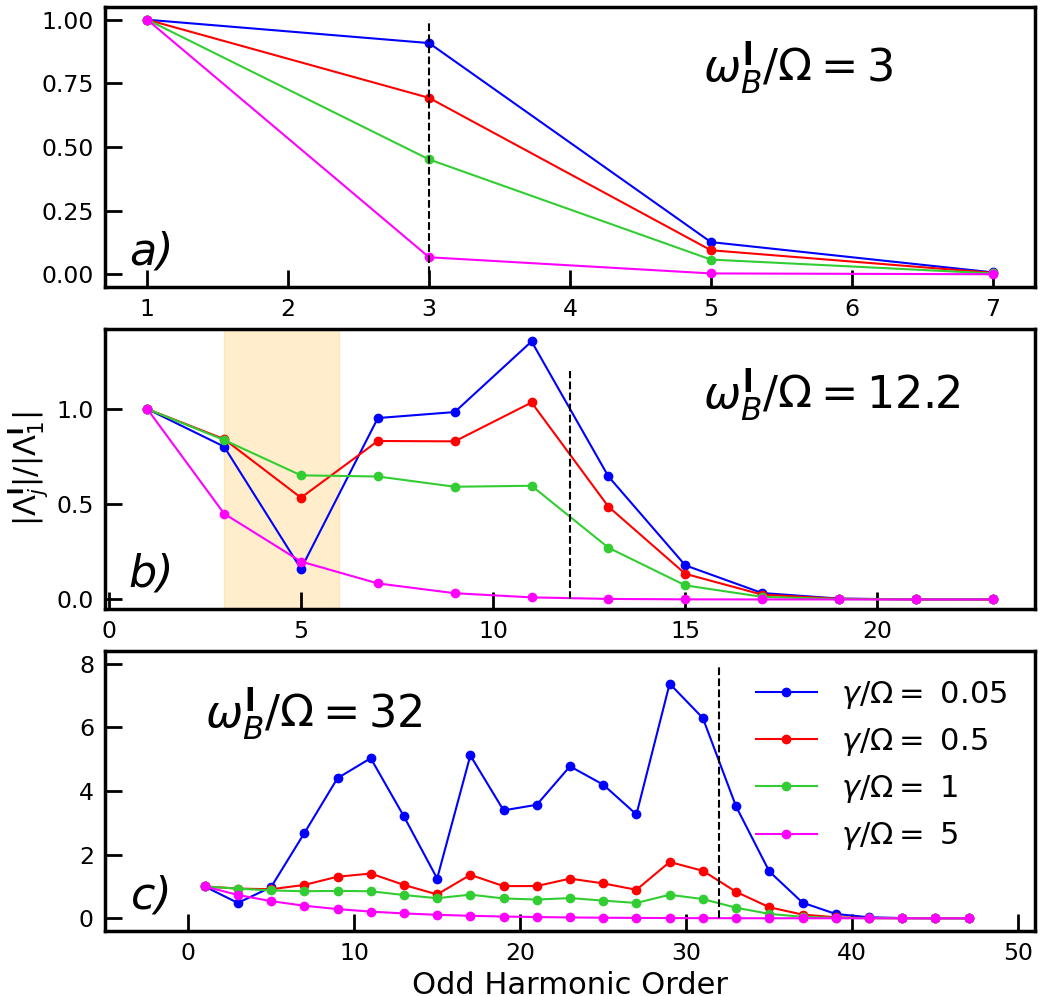}\caption{\label{fig:Odd-harmonic-coefficients} Isolated Fourier amplitudes
of the quasi-stationary state of odd harmonics divided by the amplitude
of the first harmonic, $\left|\Lambda_{j}^{\mathbf{l}}\right|/\left|\Lambda_{1}^{\mathbf{l}}\right|$,
for different dissipation strengths $\gamma/\Omega$ and a) $\omega_{B}^{\mathbf{l}}/\Omega=3$,
b) $\omega_{B}^{\mathbf{l}}/\Omega=12.2$ and c) $\omega_{B}^{\mathbf{l}}/\Omega=32.$
The black dashed line denotes the Bessel function argument $\omega_{B}^{\mathbf{l}}/\Omega$
for each case and it is seen the amplitudes decay exponentially after
that threshold. The yellow shaded line in plot b) deliniates a region
where the ratio $\left|\Lambda_{j}^{\mathbf{l}}\right|/\left|\Lambda_{1}^{\mathbf{l}}\right|$
initially increases with increasing dissipation.}
\end{figure}

In Fig. \ref{fig:Odd-harmonic-coefficients} we plot the ratio between
the odd harmonic Fourier amplitudes and the first-harmonic amplitude
$\left|\Lambda_{j}^{\mathbf{l}}\right|/\left|\Lambda_{1}^{\mathbf{l}}\right|$
of a single neighbor for a set of dissipation values $\gamma/\Omega$
and for $\omega_{B}^{\mathbf{l}}/\Omega=3,12.2$ and $32$. A first
thing to notice is how the amplitudes start dying out exponentially
for harmonics of order superior than $\omega_{B}^{\mathbf{l}}$. By
analysing Eq.(\ref{eq:coefs_mono}), one can understand this behaviour
as originating from the properties of Bessel functions. At a fixed
argument $z$, as a function of the order $n$, $J_{n}\left(z\right)$
decays exponentially as $J_{n}\left(z\right)\sim e^{-n\ln\left(z^{-1}\right)}$
for large order $n$ \citep{abramowitzHandbookMathematicalFunctions1964}.
Another interesting result is how for strong dissipation the higher
harmonics are suppressed, when compared to the first harmonic. This
result is in agreement with results from Ref.~\citep{brown_real-space_2024},
where it is interpreted as being a consequence of the suppression
of a particle's longer trajectories in real-space via dissipation,
which are responsible for hi\k{ }gh frequency contributions. However,
our results indicate that the suppression of higher harmonics with
dissipation is not monotonous with increased dissipation. As shown
in Fig.~(\ref{fig:Odd-harmonic-coefficients})b), intermediate harmonics
can have their amplitude increased with dissipation.

\begin{figure}
\includegraphics[width=1\columnwidth]{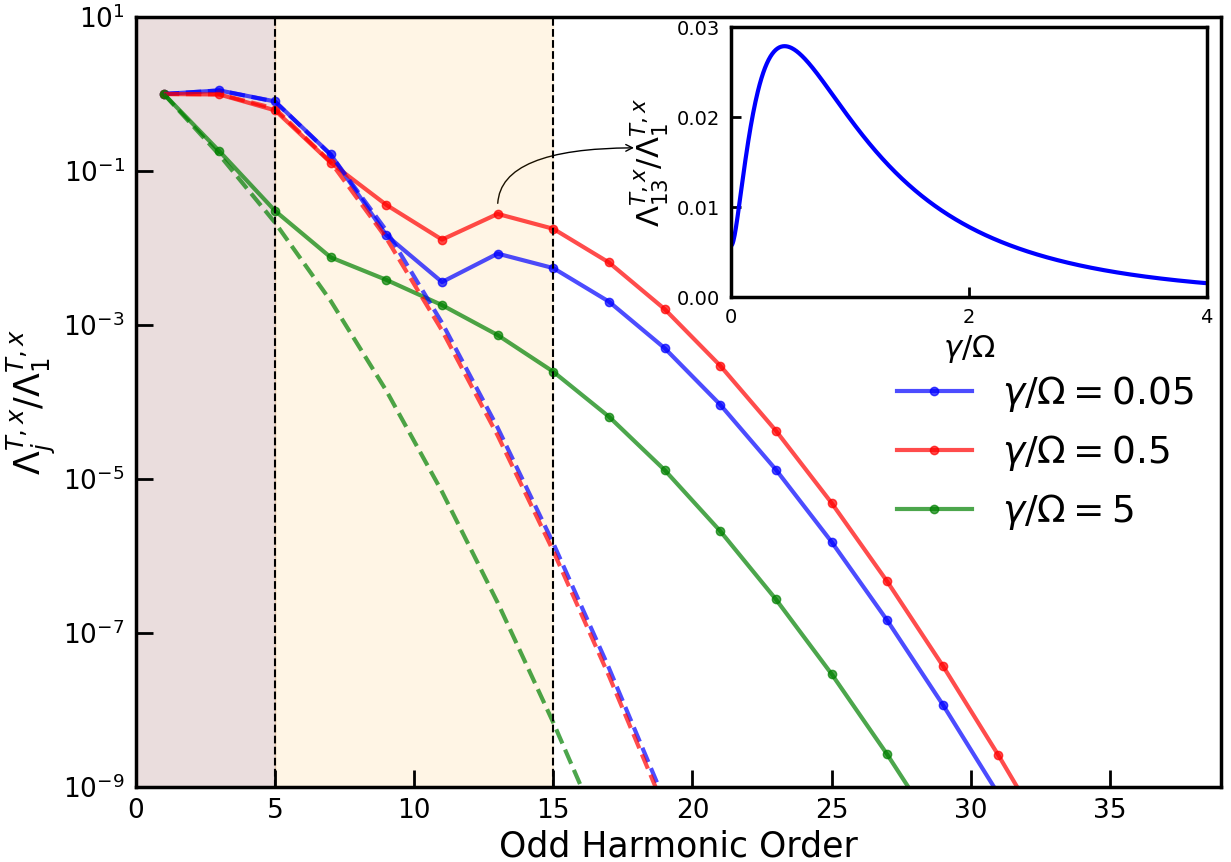}\caption{\label{fig:Odd-harmonic-2D} Odd harmonic amplitudes for a two-dimensional
nearest neighbours tight binding lattice with $l_{1}=l_{2}=1$, $t_{1}=0.2$,
$t_{2}=1$ and $\theta=1.23\,\text{rad}$, for various dissipation
coefficients in logarithmic scale. The black dashed lines show the
values of $\omega_{B}^{2}/\Omega=5$ and $\omega_{B}^{1}/\Omega=15$.
The solid lines contain the contribution from both neighbors, whereas
the dashed lines only show the contribution of the second neighbor,
which begin decaying after the threshold $\omega_{B}^{2}/\Omega=5$,
delimiting the end of the first plateau (darker shaded region). The
second plateau then develops between these two scales (lighter shaded
region). The inset shows the amplitude ratio between the 13th and
the first harmonics as a function of dissipation. It shows dissipation
can lead to a non-trivial behavior of the amplitude, which initially
increases and later decreases with respect to the first harmonic.}
\end{figure}

Having analysed the individual contributions of each hopping, we now
turn to the result obtained in the two dimensional lattice where the
two nearest-neighbour hoppings contribute to the total current. The
total amplitude contribution measured along x axis is $\Lambda_{j}^{T,x}=-2\left|C_{1}t_{1}l_{1}\Lambda_{j}^{1}+C_{2}t_{2}l_{2}\cos\theta\Lambda_{j}^{2}\right|$,
where $\Lambda_{j}^{1}$ (\textbf{$\Lambda_{j}^{2}$}) is the amplitude
contribution of the first (second) neighbor. We plot these amplitudes
in Fig. \ref{fig:Odd-harmonic-2D}, as a function of the harmonic
order, and observe the emergence of two plateaus that would not be
possible by considering only one hopping. The first plateau is observed
until one of the hopping contributions begins decaying exponentially.
Therefore, the plateau exists only until the smallest of the two scales
$\omega_{B}^{1}/\Omega$ and $\omega_{B}^{2}/\Omega$, which in this
case is the contribution from the second neighbor, $\omega_{B}^{2}/\Omega$.
The second plateau emerges between these two scales. In between these
two scales, the contributions from the second hopping start decaying
exponentially (dashed lines in Fig. \ref{fig:Odd-harmonic-2D}), however
the contributions from the first neighbor survive and give rise to
the plateau. The total amplitudes only start decaying when the harmonic
order is larger than $\omega_{B}^{2}/\Omega$. This should be visible
for any number of hoppings, given that the scales associated with
the respective $\omega_{B}^{\mathbf{l}}/\Omega$ values are well separated.
Interestingly, it is also apparent from Fig. \ref{fig:Odd-harmonic-2D}
that the behavior of the amplitudes with dissipation is highly non-perturbative.
The inset shows the amplitude ratio between the 13th and the first
harmonics as a function of dissipation. It shows dissipation can lead
to a non-perturbative behavior of the amplitude, which initially increases
and later decreases with respect to the first harmonic.

\begin{figure}
\centering{}\includegraphics[width=1\columnwidth]{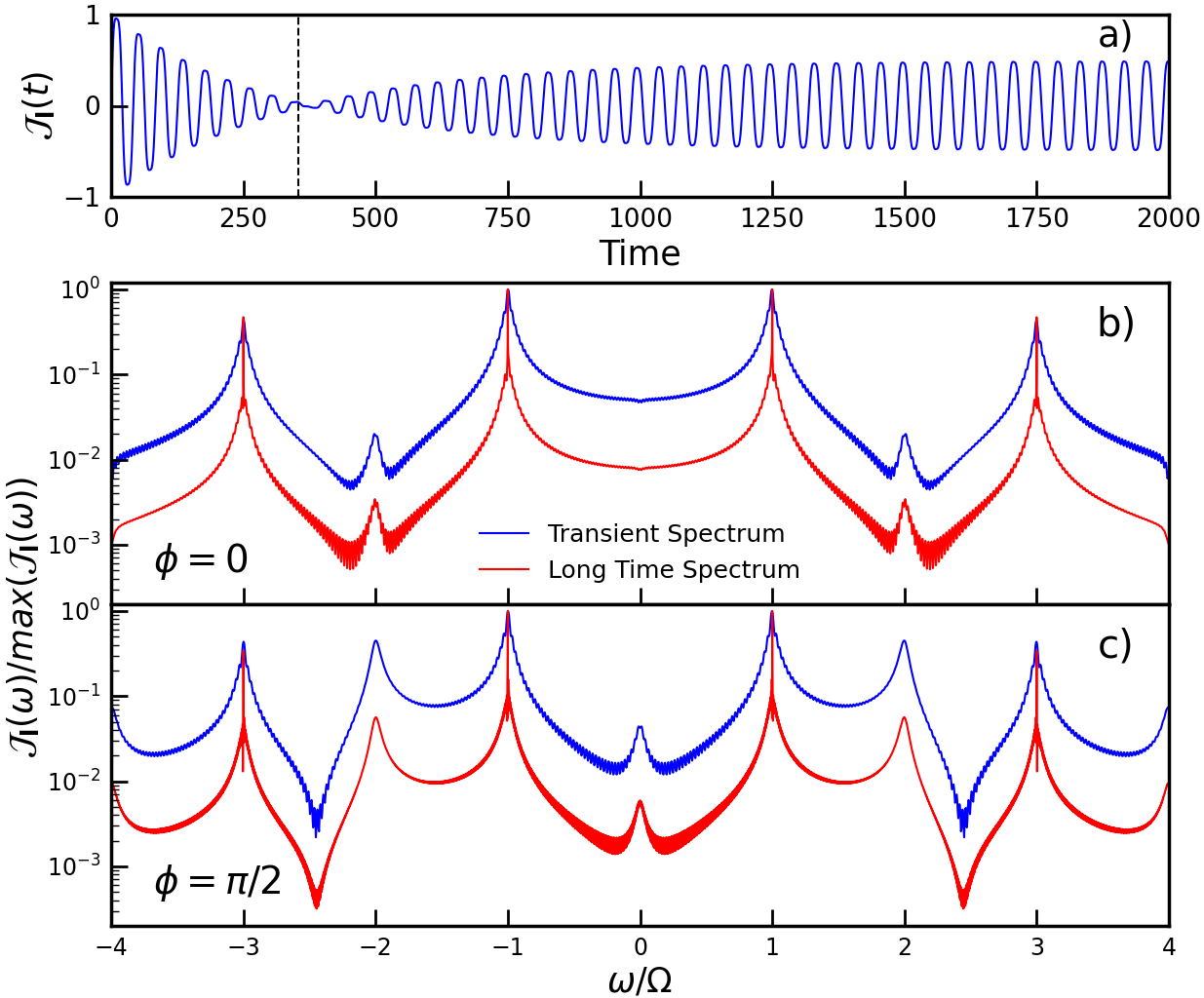}\caption{\label{fig:Transient_spec} a) Time-evolution of current for $\omega_{B}^{\mathbf{l}}/\Omega=1.54$,
$\Omega=0.15$ and $\gamma=0.003$. The black dashed line delimits
the aquisition time of for the transient regime spectrum. b) and c)
Transient and long times Fourier spectra for $\omega_{B}^{\mathbf{l}}/\Omega=2.5$,
$\Omega=0.13$ and $\gamma=0.004$ in logarithmic scale for electric
field phases b) $\phi=0$ and c) $\phi=\pi/2$. The aquisition times
for the transient and long times spectra were $2000$ and $42400$
for both plots. A small second harmonic is visible in the transient
spectrum in b), with amplitude of roughly two orders of magnitude
smaller than that of the first harmonic. . In c), however, the second
harmonic is comparable to the first harmonic in the transient spectrum
and a small DC current is also present. As the aquisition time increases,
the oscillatory long time state has a more significant contribution
and the even harmonics decrease with respect to the first harmonic
in both plots. }
\end{figure}

From Eq. \ref{eq:exact_sol_mono}, it is evident that part of the
transient regime of the current response is decomposed in even harmonics
of the electric field's frequency. In order for the even harmonics
to be detected in the transient regime, they should display many oscillations
before the transient regime fades to zero, which translates in a small
$\gamma/\Omega$ ratio. From Eqs. \ref{eq:coefs_mono}, it is easy
to see that for $\phi=0$ the even harmonic amplitudes $\chi_{2j}^{\mathbf{l},2}$
vanish for $\gamma/\Omega\ll1$ while the odd harmonic amplitudes
$\chi_{2j+1}^{\mathbf{l},1}$ approach a finite value. In practice,
this means that the even harmonics detected in the transient regime
from a Fourier spectrum of $\left\langle \mathbf{J}(t)\right\rangle $
acquired during a finite time $T_{\text{acq}}$,
\[
\mathcal{J}\left(\omega;T_{\text{acq}}\right)\equiv\int_{0}^{T_{\text{acq}}}dte^{-i\omega t}\left\langle \mathbf{J}(t)\right\rangle ,
\]
are much smaller than the odd ones. However, for $\phi\neq0$ the
even harmonics amplitude approaches a finite value in the limit $\gamma/\Omega\ll1$.In
Fig. \ref{fig:Transient_spec}, we plot the transient and long times
Fourier spectra for a system with $\omega_{B}^{\mathbf{l}}/\Omega=2.5$,
$\Omega=0.13$, $\gamma=0.004$ and $\phi=0$ (Fig. \ref{fig:Transient_spec}b)
and $\phi=\pi/2$ (Fig. \ref{fig:Transient_spec}c). The difference
between the transient and long times spectra is the aquisition time
of the signal, which is kept close or far away from the regime where
the transient current dominates, respectively. A small second harmonic
is visible in the transient spectrum for $\phi=0$, with amplitude
of roughly two orders of magnitude smaller than that of the first
harmonic. For $\phi=\pi/2$, however, the second harmonic is comparable
to the first in the transient spectrum and a small DC current is also
present. As the aquisition time increases, the oscillatory long time
state has a more significant contribution and the even harmonics decrease
with respect to the first harmonic in both plots. 

\begin{figure}
\centering{}\includegraphics[width=1\columnwidth]{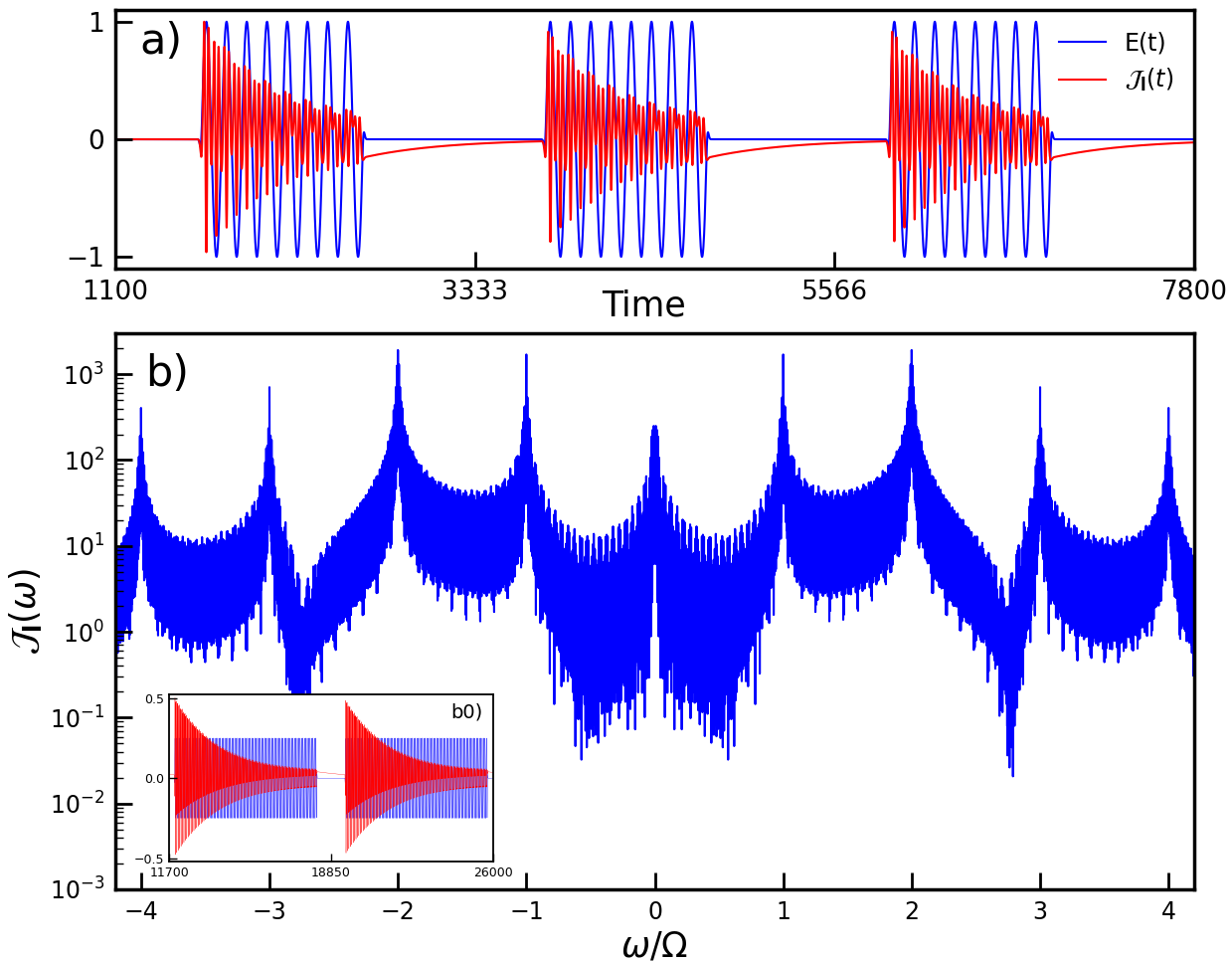}\caption{\label{fig:impulse_spectrum} a) An example of a train of square electric
pulses (solid blue line) and its corresponding current response (solid
red line) for $\omega_{B}^{\mathbf{l}}/\Omega=5$, $\Omega=0.05$
and $\gamma=0.002$. The pulse parameters are $\sigma=10$, $w=8\pi/\Omega$
and $\Delta s=34\pi/\Omega$. b) Fourier spectra for $\omega_{B}^{\mathbf{l}}/\Omega=2.5$,
$\Omega=0.1$ and $\gamma=0.0005$ in logarithmic scale. The pulse
train is composed of 12 square pulses with $\sigma=10$, $w=100\pi/\Omega$
and $\Delta s=240\pi/\Omega$. The corresponding time profile of the
current and electric field are plotted in b0). Strong even harmonics
are detected in the Fourier spectrum. }
\end{figure}

Although the emergence of even harmonics for centrossymetric systems
is bound to exist solely on the transient regime, one can think of
a setup aiming at consistently exciting the transient state, and therefore
the even harmonics, by periodically pumping the system with optical
pulses. The caveat for observing this phenomenon is that the switching
of the electric field must be faster than its carrying period, as
discussed in Appendix \ref{subsec:Even-harmonic-Vanishing}. One possible
way of realizing this condition is by using a train of square-like
pulses. The corresponding electric field can be written as 
\[
\begin{aligned}E_{\text{pulses}}(t) & =\frac{E}{4}\sin(\Omega t)\sum_{n=1}^{N_{\text{pulses}}}\left[\operatorname{erf}\!\left(\frac{t-(s_{n}-w)}{\sigma}\right)+1\right]\times\\
 & \qquad\qquad\qquad\qquad\times\left[-\operatorname{erf}\!\left(\frac{t-(s_{n}+w)}{\sigma}\right)+1\right]
\end{aligned}
\]
Where $\text{erf}(x)$ is the error function, $s_{n}$ is the center
of pulse $n$, $w$ is the pulses' half-width and $\sigma$ controls
hwo fast the pulse is switched on/off. We consider a train where the
pulses are spaced equally, i.e, $s_{n}=s_{0}+n\Delta s$ and their
spacing and width are multiples of the carrying period $T=2\pi/\Omega$
so as to respect the phase $\phi=\pi/2$ for each pulse. An example
of the pulse train and its corresponding current response are shown
in Fug. \ref{fig:impulse_spectrum}a) (in this case, the current was
obtained by numerically integrating Eq. \ref{eq:exact_sol_general}).
The system is pumped out of equilibrium into the transient state by
the first pulse during its width. After the pulse is switched off,
the system is allowed to relax back into the initial state before
being excited again by the upcoming pulse, allowing the excitation
of the same transient state. Therefore, the pulses' spacing, width
and carrying frequency can be tuned for a given dissipation strength
to allow the consistent emergence of strong even harmonics. Fig. \ref{fig:impulse_spectrum}b)
depicts the Fourier spectrum (and its corresponding time profile in
Fig. \ref{fig:impulse_spectrum}b0) ) of the current response for
a given impulse, showcasing a second and fourth harmonics of the same
orders of magnitude as the first and third, respectively. 

\subsection{Constant field}

\begin{figure}
\includegraphics[width=1\columnwidth]{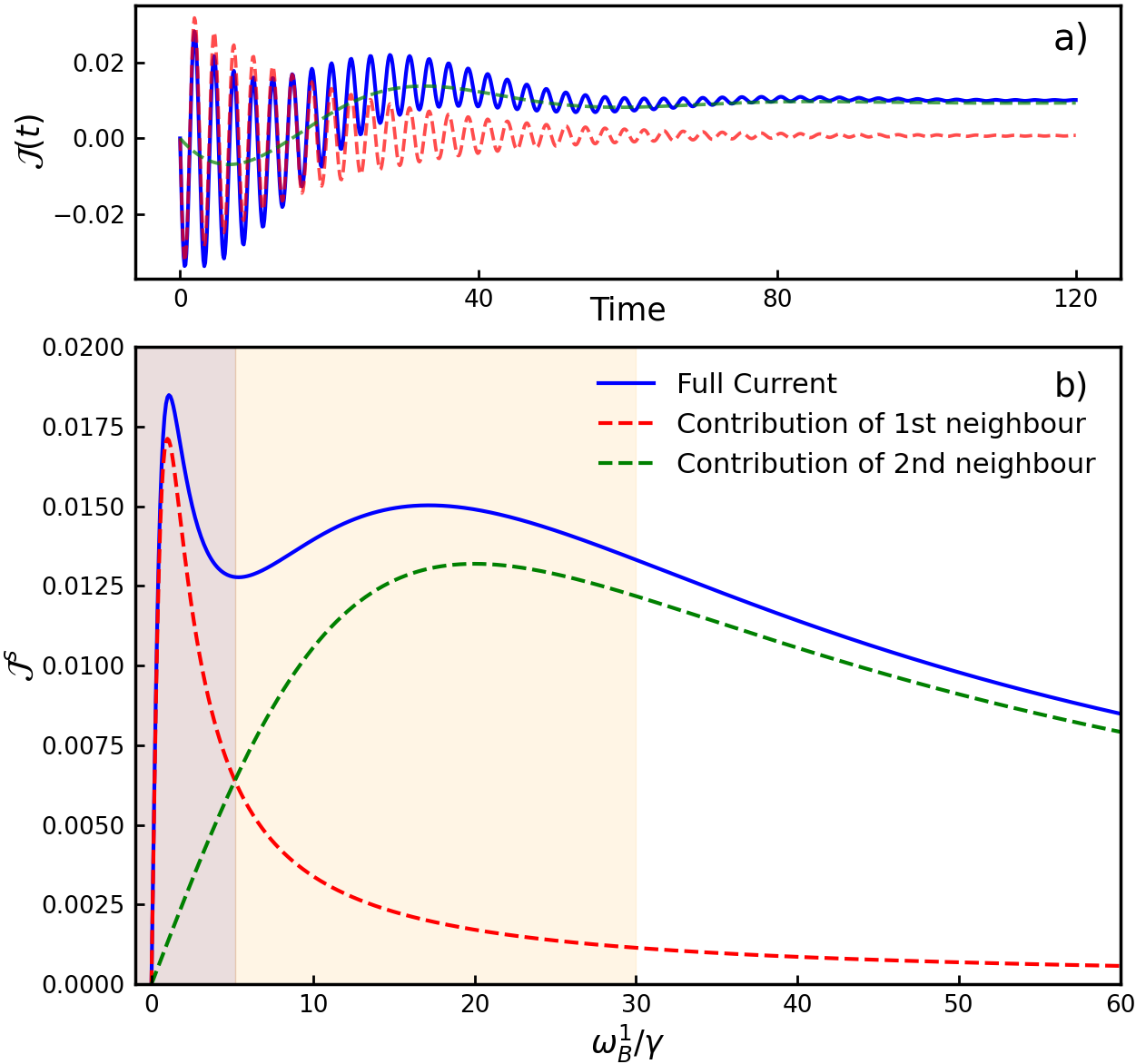}\caption{\label{fig:esaki_tsu_curves} a) Current time evolution of a 2D nearest
neighbor tight-binding lattice subject to a constant electric field
with $l_{1}=l_{2}=1$, $t_{1}=0.3$, $t_{2}=0.7$ and $\theta=1.52$
rads. The solid line is the full solution, whereas the red and green
dashed lines are the contributions of the first and second neighbor,
respectively. It is seen that the inital transient regime displaying
Bloch Oscillations decays to a constant, stationary value. b) Stationary
current of same 2D system as a function of $\omega_{B}^{1}/\gamma$.
The solid line is the full stationary current, while the dashed lines
are the contributions of each of the neighbors. Through the interplay
of both neighbor contributions, two maximums (darker and lighter shaded
areas) are achieved.}
\end{figure}

In Fig. \ref{fig:esaki_tsu_curves} we plot the time evolution of
the current in a 2D lattice subject to a constant electric field (Fig.
\ref{fig:esaki_tsu_curves} a) and the total stationary current profile
of the same 2D system as a function of $\omega_{B}^{1}/\gamma$ (Fig.
\ref{fig:esaki_tsu_curves} b). The time evolution plot shows the
initial transient regime displaying Bloch Oscillations decaying exponentially
to a constant, finite value. The stationary current profile of each
neighbor contribution is also plotted in Fig. \ref{fig:esaki_tsu_curves}
b. For small $\omega_{B}^{\mathbf{l}}/\gamma$ the stationary current
goes to zero. This is because strong dissipation eventually suppresses
propagation of charge carriers inside the sample. For large $\omega_{B}^{\mathbf{l}}/\gamma$
ratio, the stationary current also vanishes. This corresponds to the
limit where there is no dissipation, and therefore Bloch Oscillations
become persistent and no stationary current is formed.

An intuitive picture to understand this behaviour can be drawn in
analogy to the Drude argument: Suppose we have a particle roaming
in a disordered crystal subject to a constant electric field and that,
like the original Drude argument, everytime it colides with an impurity,
its momentum resets to zero. If no collisions between the particle
and the impurities occur, the particle will only oscillate in place,
displaying BOs, and therefore no DC current is generated. However,
if sometime during its oscillation the particle collides with an impurity,
the oscillation restarts at that point in space, and the particle
will now oscillate in turn of that point, shifted from the place it
started. If the collisions occur somewhat regularly, the particle
will will be able to drift along the crystal, motivating the increase
of the current seen in the Esaki-Tsu relation by increasing dissipation.
As such, we can see the Esaki-Tsu current as a remnant of BOs, which
are frustrated by the presence of dissipation.

The generalization of the Esaki-Tsu relation to any single-band model
accommodates extra effects that come from the interplay between the
different hopping contributions. In Fig. \ref{fig:esaki_tsu_curves}
b), the large separation between the scales $\omega_{B}^{1}/\gamma$
and $\omega_{B}^{2}/\gamma$ allows for a stationary current that
displays two maximum values, instead of the one characteristic of
the Esaki-Tsu equation.

\section{Comparison with Perturbation Theory\label{sec:Comparison-with-Perturbation}}

In this section, we aim at comparing our exact result for the current
produced by a constant field to the result obtained from perturbation
theory. For illustrative purposes, we apply the general time-resolved
solution for a constant field (Eq. \ref{eq:const_field_sol})to a
simple 1D tight-binding chain. Starting with the case of no dissipation,
$\gamma=0$, the average current displays the famous Bloch Oscillations,
\[
J^{\gamma=0}\left(t\right)=-2ta\frac{e}{\hbar}C\sin\left(\omega_{B}t\right),
\]
where $\omega_{B}=\frac{e}{\hbar}Ea$ is the Bloch frequency and $C=\int_{-\pi/a}^{\pi/a}dkf(\varepsilon_{k})e^{ika}=\int_{-\pi/a}^{\pi/a}dkf(\varepsilon_{k})\cos ka$.
Notice that, with no dissipation, these oscillations do not decay
to a constant value. The problem of trying to apply a peturbation
scheme in this particular case lies in the fact that the electric
field is coupled to the time. In other words, the gauge potential
grows linearly in time $A(t)\propto Et$ and therefore the expansion
parameter of the sinosoidal function $\omega_{B}t$ cannot be regarded
as small for any time instance. Consequently, perturbation theory
cannot be applied equally across all time, and is only valid for times
smaller than a critical time $t_{c}$ defined by $\omega_{B}t_{c}\ll1$.
For $t>t_{c}$, perturbation theory fails. 

On the other hand, if one is to add a finite dissipation rate $\gamma$,
the current will display these Bloch oscillations decaying to a constant
value given by the Esaki-Tsu relation 
\begin{equation}
\begin{aligned}J\left(t\right) & =2ta\frac{e}{\hbar}C\frac{\omega_{B}/\gamma}{1+\left(\omega_{B}/\gamma\right)^{2}}\left\{ -1+\right.\\
 & \qquad\qquad+\left.e^{-\gamma t}\left(\frac{\omega_{B}}{\gamma}\sin\left(\omega_{B}t\right)+\cos\left(\omega_{B}t\right)\right)\right\} 
\end{aligned}
\label{eq:exact_current_1D}
\end{equation}

The drastic difference from the previous case is the emergence of
another parameter $\omega_{B}/\gamma$ which can be regarded as small
for all times, and for which perturbation theory can be applied. For
example, a direct expansion of this expression up to linear order
in $\omega_{B}/\gamma$ gives 
\begin{equation}
\begin{aligned}J^{(1)}\left(t\right)=2ta\frac{e}{\hbar}C\frac{\omega_{B}}{\gamma}\left\{ e^{-\gamma t}\cos\left(\omega_{B}t\right)-1\right\} .\end{aligned}
\label{eq:exact_first_order}
\end{equation}
Its worthy to reiterate that here, we have performed a Taylor expansion
on the ratio $\omega_{B}/\gamma$ and not solely on the electric field,
which is why we retain all orders of the electric field on the cosine
function $\cos\left(\omega_{B}t\right)$.

We now compare this equation to the result of the current from the
Kubo formula. The Kubo formula for the first-order current response
of a 1D tight binding chain with dissipation obeying Eq. \ref{eq:QDM_many_scatts}
subject to a monochromatic electric field reads (See Appendix \ref{sec:Perturbation-Theory}
for details)
\begin{equation}
\begin{aligned}J^{\left(1\right)}\left(\omega\right)=-i & \frac{E\left(\omega\right)}{\left(\omega+i\gamma\right)}\int_{-\pi/a}^{\pi/a}dk\frac{df\left(\varepsilon_{k}\right)}{d\varepsilon}v^{2}\left(k\right)\end{aligned}
.
\end{equation}
For the 1D tight binding chain $v\left(k\right)=-2ta\frac{e}{\hbar}\sin(ka)$.
By inverting the dispersion relation $k=\arccos\left(-\varepsilon/2t\right)$,
writing $\frac{df\left(\varepsilon_{k}\right)}{d\varepsilon}=\frac{df\left(\varepsilon_{k}\right)}{d\varepsilon}\frac{dk}{d\varepsilon}$
and performing an integration by parts, the current may be expressed
as 
\begin{equation}
\begin{aligned}J^{\left(1\right)}\left(\omega\right)=-i & D\frac{E\left(\omega\right)}{\left(\omega+i\gamma\right)}\end{aligned}
,\label{eq:Drude}
\end{equation}
with $D=2ta^{2}\frac{e^{2}}{\hbar^{2}}\int_{-\pi/a}^{\pi/a}dkf\left(\varepsilon_{k}\right)\cos k=2ta^{2}\frac{e^{2}}{\hbar^{2}}C$.
The optical conductivity is $\begin{aligned}\sigma\left(\omega\right)= & \frac{i}{\hbar}D/\left(\omega+i\gamma\right)\end{aligned}
$, where $D$ is the Drude weight and the DC conductivity is $\sigma_{DC}=-iD/\gamma$.
If we transform Eq.~(\ref{eq:Drude}) back to time-domain, we get
\begin{equation}
\begin{aligned}J^{\left(1\right)}\left(t\right) & =-DE\frac{1-e^{-\gamma t}}{\gamma}.\end{aligned}
\label{eq:first_order_const}
\end{equation}
Directly comparing this expression to Eq. \ref{eq:exact_first_order}
shows that the only difference lies solely on the transient term:
While the former retains the complete time depedence of the cosine,
the one derived from the Kubo formula only keeps the first term of
the cosine with respect to $\omega_{B}t$, $\cos\omega_{B}t=1+O\left(E^{2}\right)$.
As for the case of no dissipation, this comes from the fact that the
Kubo is expanding the current on powers of the gauge potential $A(t)\propto Et$,
which means also expanding the cosine function to first order in the
gauge potential. As it was shown, this is bound to fail for times
larger than $1/\omega_{B}$. This results in the Kubo formula incorrectly
predicting the first order of the transient regime for every instance
of time. 

Interestingly enough, however, it gives the correct result for the
first order of the stationary current. This result is due to the fact
that the underlying assumption of the Kubo formula that $A\left(t\right)\ll1$
inherently results in $E\ll1$ for the constant field. Therefore,
it also captures the first order of the stationary current on the
electric field. Because $\omega_{B}/\gamma\propto E$, this coincides
with the first order of the exact stationary current given by Eq.
\ref{eq:exact_current_1D} on $\omega_{B}/\gamma$. This shows that,
as long as $\omega_{b}/\gamma\ll1$, a perturbation theory treatment
of the current response can give the right result for the stationary
current and the Drude weight obtained from the Kubo formula can be
safely used to characterize the conductivity of the material in the
stationary regime under a constant electric field. Conversely, it
is bound to fail for the transient current, although it gets exponentially
supressed. 

\section{Conclusion\label{sec:Conclusion}}

In conclusion, we have sucessfully obtained an exact, time-resolved
analytical solution for the current response of a driven dissipative
single-band tight binding system evolving according to the RTA and
under an uniform electric field with generic time-dependence. We have
applied the solution for the two limiting cases of a monochromatic
field and a constant field and obtained analytical expressions for
both cases. The monochromatic driven system generally displays both
even and odd transient harmonics that get exponentially supressed
and give rise to a stable, oscillatory state composed solely of odd
harmonics, respecting the inherent inversion symmetry of the single
band system. We have concluded that dissipation is a fundamental ingredient
for this relation, since for the non-dissipative case, the current
may display persistent even harmonics if the field is suddenly switched
fast enough. The stable odd harmonics were seen to form plateau regions
with well defined boundaries, after which they decay exponentially.
Furthermore, the behavior of the stable harmonics with dissipation
was also analytically obtained and their non-perturbative tendency
was under2lined, with some harmonics initially increasing with dissipation
with respect to the driving frequency. As for the constant field case,
an analytical expression showcasing decaying Bloch oscillations towards
a stationary state was derived. The stationary state was seen to be
a generalization of the Esaki-Tsu relation for the generic single-band
tight-binding system. Finally, we have further compared our exact
result for the constant field to that obtained by the usual perturbation
theory framework, and showed that while PT cannot correctly predict
the transient regime response across all time, it does correctly perturbatively
describe the stationary regime nonetheless. We are hopeful that this
results shine some light on the effects of dissipation on driven systems
and the validity of perturbation theory for this class of problems. 

\section{Appendix}

\subsection{Even harmonic Vanishing with Adiabatic Switching\label{subsec:Even-harmonic-Vanishing}}

Let us consider a one dimensional, nearest-neighbour tight binding
system perturbed by an electric field of the form $E\left(t\right)=E\Theta\left(t\right)\cos\left(\Omega t+\phi\right)$
suddenly switched on at $t=0$, whose exact solution is given by Eq.
\ref{eq:exact_sol_mono}. By performing a perturbative expansion to
first order in the electric field to Eq. \ref{eq:exact_sol_mono}
for zero dissipation, we get that the current is 
\[
\mathcal{J}\left(t\right)=-\frac{2tCe^{2}}{\hbar^{2}}\frac{E}{\Omega}\left\{ \frac{1}{2}\sin\left(\left(\Omega t-t_{0}+\phi\right)\right)-\sin\left(\phi\right)\right\} +O\left(E^{2}\right).
\]
The linear contribution to the current displays not only a contribution
to the first harmonic, but a contribution to the DC current as well,
which is thought to be prohibited in centrossymmetric systems. The
other even harmonics will also arise as finite contributions originating
from higher odd powers on the electric field. To gauge the effect
of adiabatic switching on even harmonics, we will now consider an
electric field of the form $E\left(t\right)=Ef\left(t\right)\cos\left(\Omega t+\phi\right)$,
where $f\left(t\right)$ is the function describing the switching
of the electric field. The linear contribution to the current can
be written, in Fourier space, as 
\[
\mathcal{J}\left(\omega\right)=\sigma\left(\omega\right)E\left(\omega\right)+O\left(E^{2}\right),
\]
Where $\sigma\left(\omega\right)=-\text{\ensuremath{\lim_{\eta\rightarrow0}}}\frac{eti}{\hbar}C/\left(\omega+i\eta\right)$
is the optical conductivity and $E\left(\omega\right)=\lim_{\eta\rightarrow0}\int_{-\infty}^{\infty}Ef\left(t\right)e^{i\omega t-\eta\left|t\right|}\cos\left(\Omega t+\phi\right)$
and $\mathcal{J}\left(\omega\right)$ the Fourier transforms of the
electric field and the current, respectively. To simplify calculations,
let us change the time frame by setting the electric field to $E\left(t\right)=Ef\left(t-t_{0}\right)\cos\left(\Omega\left(t-t_{0}\right)+\phi\right)$
and let us assume that at time $t=0$ the electric field is zero.
Through integration by parts, the Fourier transform of the electric
field can be written as
\[
\begin{aligned}E\left(\omega\right) & =-\frac{1}{2}\lim_{\eta\rightarrow0}\int dte^{i\omega t-\eta t}\frac{df(t-t_{0})}{dt}\times.\\
 & \quad\quad\quad\quad\quad\times\left(\frac{e^{i\left(\Omega\left(t-t_{0}\right)+\phi\right)}}{i(\omega+\Omega)-\eta}+\frac{e^{-i\left(\Omega\left(t-t_{0}\right)+\phi\right)}}{i(\omega-\Omega)-\eta}\right)
\end{aligned}
\]
The first order contribution to the DC current $\mathcal{J}\left(0\right)$
is therefore given by
\begin{equation}
\begin{aligned}\mathcal{J}^{\left(1\right)}\left(0\right) & =\lim_{\eta\rightarrow0}\frac{te^{2}CE}{\hbar^{2}\left(\Omega^{2}+\eta^{2}\right)}\int dt\frac{df(t-t_{0})}{dt}e^{-\eta t}\times.\\
 & \times\left(\left(\frac{\Omega}{\eta}\sin\left(\Omega\left(t-t_{0}\right)+\phi\right)-\cos\left(\Omega\left(t-t_{0}\right)+\phi\right)\right)\right)
\end{aligned}
\label{eq:DC_current_firstorder}
\end{equation}
The DC current contribution from the linear term is now written in
terms of the derivative of the switching function $f\left(t\right)$.
For very fast switchings of the electric field, i.e, faster than its
period, the integral in Eq. \ref{eq:DC_current_firstorder} will be
dominated by the derivative's peak at instant $t_{0}$ and the DC
current will have a finite contribution. For slower switchings (larger
than its period), $df/dt$ gets broader and the contributions from
the oscillating functions in Eq. \ref{eq:DC_current_firstorder} become
more relevant. If sufficient oscillations are included, the integral
of these functions start averaging closer to zero as $\frac{df(t-t_{0})}{dt}$
becomes broader. Therefore, adiabatic switching destroy direct current.
The other even harmonics have an analogous behaviour.

\begin{figure}
\includegraphics[width=1\columnwidth]{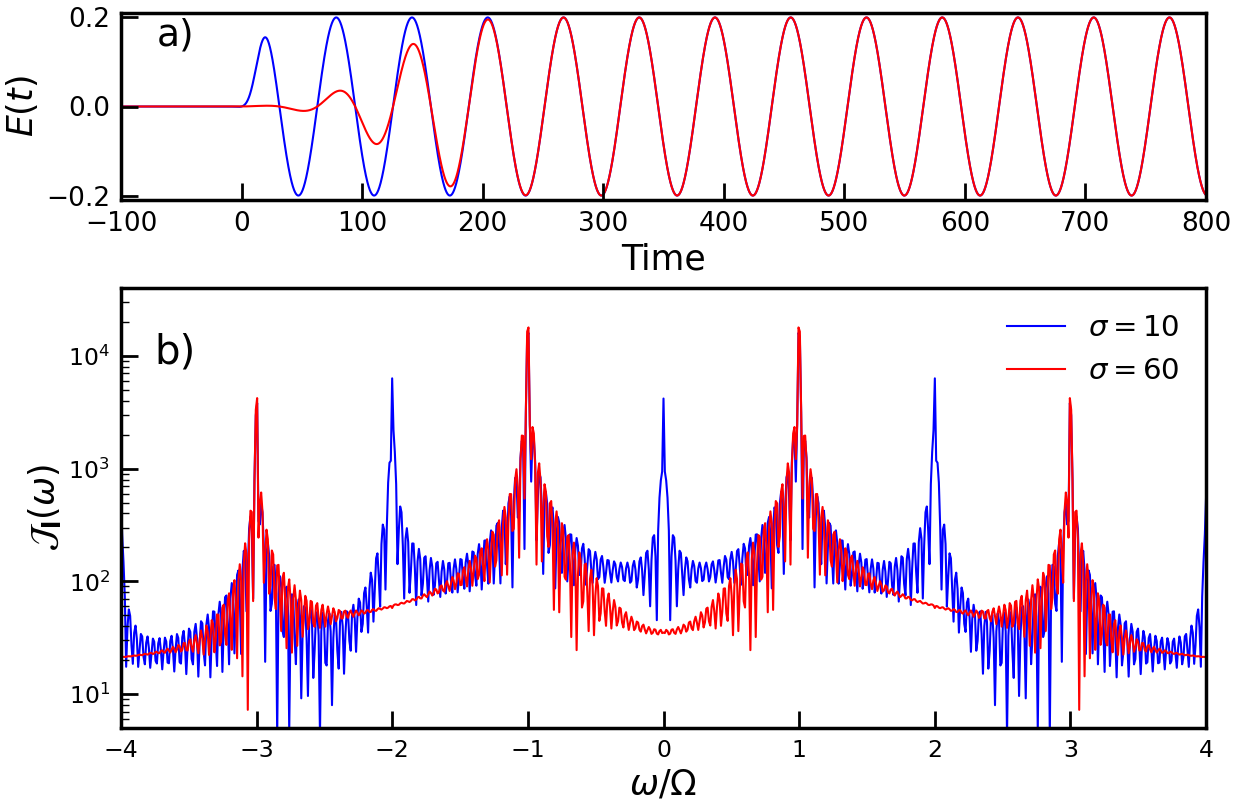}\caption{\label{fig:even_harm_adiabatic} a) Electric field time dependence
for $\phi=\pi/2$ for two different switchings with $\sigma=10$ and
$\sigma=60$ and b) Fourier spectra of the currents generated by those
electric fields respectively. We see the emergence of even harmonics
if the electric field is turned on faster than a single cycle of the
electric field, but they quickly vanish as the electric field is made
to turn on slower than a couple cycles.}
\end{figure}

In Fig. \ref{fig:even_harm_adiabatic}, we plot the time dependence
of two adiabatically switched electric fields with $\phi=\pi/2$ and
$f\left(t\right)=\frac{1}{2}\left(\text{erf}\left(\left(t-t_{0}\right)/\sigma\right)+1\right)$,
with $\text{erf}\left(t\right)$ being the error function, and their
respective current's Fourier spectra. It shows that for the fast switching
electric field, whose switching has a duration smaller than its period,
the even harmonics are present. Conversely, for the slow switching
electric field, that switches through a duration of roughly 3 periods,
the even harmonics vanish.

\subsection{Perturbation Theory\label{sec:Perturbation-Theory}}

Here we derive the first order current response for generic crystalline
system with dissipation. The Hamiltonian at time $t$ is block diagonal
in the Bloch basis through matrices $\bar{H}\left(\boldsymbol{k},\boldsymbol{A}(t)\right)$
whose dimensions equal the number of internal degrees of freedom of
the system. The Hamiltonian is written in terms of this matrices as
\[
\hat{H}(t)=\sum_{\boldsymbol{k},\sigma\sigma'}\bar{H}_{\sigma\sigma'}\left(\boldsymbol{k},\boldsymbol{A}(t)\right)\ket{\sigma}\bra{\sigma'}\otimes\ket{\boldsymbol{k}}\bra{\boldsymbol{k}},
\]
where $\sigma$ index the internal degrees of freedom of the system
and $\bar{H}_{\sigma\sigma'}\left(\boldsymbol{k},\boldsymbol{A}(t)\right)\equiv\bra{\sigma}\bar{H}\left(\boldsymbol{k},\boldsymbol{A}(t)\right)\ket{\sigma'}$.
Moreover, every operator appearing in this derivation will also be
block diagonal in $\mathbf{k}$. To simplify notation, we will from
now on work the matrices $\bar{H}$ and their analogs for the other
operators whose dependency on $\mathbf{k}$ is implied, knowing that
we can obtain the full operator by integrating in $\mathbf{k}$. We
firstly write the Hamiltonian as $\bar{H}(t)=\bar{H}_{0}+\hat{V}(t)$,
where $\bar{H}_{0}$ is the Hamiltonian for the system with no gauge
field and $\hat{V}(t)$ contains all of the dependency of said field.
We can calculate the current operator as a functional derivative of
the Hamiltonian with respect to the gauge potential $\hat{J}_{\alpha}\left(t\right)=-\frac{1}{\mathcal{V}}\frac{\partial\bar{H}(t)}{\partial A_{\alpha}}$.
For what follows, we will only require the expansion of this operator
up to first order in the gauge potential, which requires the expansion
of the Hamiltonian up to second order. Doing so yields:
\[
\begin{aligned}\hat{V}(t) & =\sum_{\lambda}\left\{ \frac{e}{\hbar}v_{\alpha}^{\lambda\lambda'}A_{\alpha}(t)+\frac{1}{2}\Delta_{\alpha\beta}^{\lambda\lambda'}A_{\alpha}(t)A_{\beta}(t)+...\right\} \ket{\lambda}\bra{\lambda'},\\
\hat{J}_{\alpha}(t) & =\sum_{\lambda\lambda'}\left\{ v_{\alpha}^{\lambda\lambda'}+\Delta_{\alpha\beta}^{\lambda\lambda'}A_{\beta}(t)+...\right\} \ket{\lambda}\bra{\lambda'},
\end{aligned}
\]
Where $\lambda$ indexes the bands of the system and we defined
\[
\begin{aligned}v_{\alpha}^{\lambda\lambda'} & =\sum_{\sigma\sigma'}\left.\frac{\partial\bar{H}_{\sigma\sigma'}}{\partial A_{\alpha}}\right|_{\mathbf{A}=\mathbf{A}\left(t\right)}\braket{\lambda}{\sigma}\braket{\sigma'}{\lambda'},\\
\Delta_{\alpha\beta}^{\lambda\lambda'} & =\frac{e^{2}}{\hbar^{2}}\sum_{\sigma\sigma'}\left.\frac{\partial^{2}\bar{H}_{\sigma\sigma'}}{\partial A_{\alpha}\partial A_{\beta}}\right|_{\mathbf{A}=\mathbf{A}\left(t\right)}\braket{\lambda}{\sigma}\braket{\sigma'}{\lambda'}.
\end{aligned}
\]
The operators are called the velocity operator and the diamagnetic
tensor, respectively. The equation we want to solve, already diagonalized
in Bloch basis, is
\begin{equation}
\frac{d\rho(t)}{dt}=-\frac{i}{\hbar}\left[\bar{H}\left(t\right),\rho\left(t\right)\right]-\gamma\left(\rho(t)-\rho^{\text{eq}}(t)\right),\label{eq:DM_crystal}
\end{equation}
Where $\rho(t)$, $\bar{H}(t)$ and $\rho^{\text{eq}}(t)$ are matrices
of dimensional equal to the number of bands of the system. We assume
an expansion of the form $\rho(t)=\rho^{\left(0\right)}(t)+\rho^{\left(1\right)}(t)+\rho^{\left(2\right)}(t)+...$
(where $\rho^{\left(n\right)}(t)\sim\left|\mathbf{A}\left(t\right)\right|^{n}$)
is possible. Because $\rho_{\mathbf{}}^{\text{eq}}(t)$ also depends
on the gauge potential, we need to expand it as $\rho_{\mathbf{}}^{\text{eq}}(t)=\rho^{\text{eq},\left(0\right)}(t)+\rho^{\text{eq},\left(1\right)}(t)...$
in order to apply perturbation theory. The orders can be calculated
by expanding the Fermi-Dirac function in a Matsubara sum (see Appendix
\ref{subsec:Matsubara-Calculation-for}) and give
\[
\begin{alignedat}{1}\rho_{\lambda\lambda'}^{\text{eq}\left(0\right)}\left(t\right) & =f_{\lambda}\delta_{\lambda\lambda'},\\
\rho_{\lambda\lambda'}^{\text{eq}\left(1\right)}\left(t\right) & =\begin{cases}
\left(\frac{f_{\lambda}-f_{\lambda'}}{\varepsilon_{\lambda}-\varepsilon_{\lambda'}}\right)V_{\lambda\lambda'}\left(t\right), & \varepsilon_{\lambda}\neq\varepsilon_{\lambda'},\\
\frac{df_{\lambda}}{d\varepsilon}V_{\lambda\lambda'}\left(t\right), & \varepsilon_{\lambda}=\varepsilon_{\lambda'}.
\end{cases}
\end{alignedat}
\]
Where we defined $f_{\lambda}\equiv f\left(\varepsilon_{\lambda}\right)$
and $\varepsilon_{\lambda}\equiv\varepsilon_{\lambda}(\mathbf{k})$
is the dispersion relation of band $\lambda$ evaluated at $\mathbf{k}$.
One only needs to isolate the different orders of the gauge potential
in Eq. \ref{eq:DM_crystal} and iteratively solve for each order.
Starting with the zeroth order, it is not hard to show that $\rho_{\lambda,\lambda'}^{\left(0\right)}(t)=f_{\lambda}\delta_{\lambda\lambda'},$
since the initial condition is $\rho_{\lambda,\lambda'}^{\left(0\right)}(0)=\rho_{\lambda,\lambda'}^{\text{eq},\left(0\right)}=f_{\lambda}\delta_{\lambda\lambda'}$.
Then, after performing the Fourier transform, the first order correction
obeys the equation
\begin{equation}
\begin{aligned}\left(i\omega+i\omega_{\lambda\lambda'}+\gamma\right)\rho_{\lambda,\lambda'}^{\left(1\right)}(\omega)=\frac{i}{\hbar}\left[f_{\lambda}-f_{\lambda'}\right]V_{\lambda\lambda'}\left(\omega\right)\\
+\gamma\rho_{\lambda\lambda'}^{\text{eq}\left(1\right)}\left(\omega\right),
\end{aligned}
\label{eq:kubo_midpoint}
\end{equation}
Where $\omega_{\lambda\lambda'}=\left(\varepsilon_{\lambda}-\varepsilon_{\lambda'}\right)/\hbar$.
At this point it is convenient to separate the cases where the two
considered energies are degenerate or not. Eq. \ref{eq:kubo_midpoint}
then yields
\[
\rho_{\lambda,\lambda'}^{\left(1\right)}(\omega)=\left\{ \begin{array}{cc}
\frac{\gamma}{\left(i\omega+\gamma\right)}\frac{df_{\lambda}}{d\varepsilon}V_{\lambda\lambda'}\left(\omega\right) & ,\varepsilon_{\lambda}=\varepsilon_{\lambda'},\\
\left\{ i+\frac{\gamma}{\omega_{\lambda\lambda'}}\right\} \frac{f_{\lambda}-f_{\lambda'}}{\hbar\left(i\omega+i\omega_{\lambda\lambda'}+\gamma\right)}V_{\lambda\lambda'}\left(\omega\right) & ,\varepsilon_{\lambda}\neq\varepsilon_{\lambda'}.
\end{array}\right.
\]
The first order correction to the current will be
\[
\begin{aligned}\left\langle \mathcal{J}_{i}^{\left(1\right)}\left(\omega\right)\right\rangle  & =\text{Tr}\left[\rho^{\left(1\right)}\left(\omega\right)v_{i}\right]\\
 & +\text{Tr}\left[\rho^{\left(0\right)}\left(\omega\right)\Delta_{ij}\right]A_{j}(\omega)
\end{aligned}
\]
The calculation of the first term is straightforward. To calculate
the second term, we resort to static perturbation theory to a time-independent
gauge potential $\mathbf{A}$. 

We consider that $A$ is a time-independent parameter. The average
current for such a system, to first order in $A$, has a very similar
form to the dynamic quantity above:
\begin{equation}
\begin{aligned}\left\langle \mathcal{J}_{i}^{A}\right\rangle  & =\text{Tr}\left(\rho^{eq\left(0\right)}\Delta^{ij}\right)A_{j}+e\text{Tr}\left(\rho^{eq\left(1\right)}v_{i}\right)+\mathcal{O}(\mathbf{A}^{2})\end{aligned}
,\label{eq:av_current_static}
\end{equation}
Where $\rho^{eq\left(0\right)}$ and $\rho^{eq\left(1\right)}$ are
the quantities calculated in \ref{eq:rho_eq} (but without the time
dependence). The second term is, therefore 
\[
\begin{aligned}\text{Tr}\left(\rho^{eq\left(1\right)}v_{i}\right)=e\sum_{\left\{ \varepsilon_{\alpha}\neq\varepsilon_{\beta}\right\} }\left(\frac{f\left(\varepsilon_{\alpha}\right)-f\left(\varepsilon_{\beta}\right)}{\varepsilon_{\alpha}-\varepsilon_{\beta}}\right)v_{j}^{\alpha\beta}v_{i}^{\beta\alpha}A_{j}\\
+e\sum_{\left\{ \varepsilon_{\alpha}=\varepsilon_{\beta}\right\} }\frac{df\left(\varepsilon_{\alpha}\right)}{d\varepsilon}v_{j}^{\alpha\beta}v_{i}^{\beta\alpha}A_{j}.
\end{aligned}
\]
Therefore, derivating the equation \ref{eq:av_current_static} in
order to $A_{j}$ and setting it to zero gives the expression for
the contribution of the diamagnetic term:

\[
\begin{aligned}\text{Tr}\left(\rho^{eq\left(0\right)}\Delta^{ij}\right)=\left.\frac{d\left\langle \mathcal{J}_{i}^{A}\right\rangle }{dA_{j}}\right|_{\mathbf{A}=0}-\sum_{\left\{ \varepsilon_{\alpha}=\varepsilon_{\beta}\right\} }\frac{df\left(\varepsilon_{\alpha}\right)}{d\varepsilon}v_{j}^{\alpha\beta}v_{i}^{\beta\alpha}\\
-\sum_{\left\{ \varepsilon_{\alpha}\neq\varepsilon_{\beta}\right\} }\left(\frac{f\left(\varepsilon_{\alpha}\right)-f\left(\varepsilon_{\beta}\right)}{\varepsilon_{\alpha}-\varepsilon_{\beta}}\right)v_{j}^{\alpha\beta}v_{i}^{\beta\alpha}.
\end{aligned}
\]
The term $\left.\frac{d\left\langle \mathcal{J}_{i}^{A}\right\rangle }{dA_{j}}\right|_{\mathbf{A}=0}$
is pertaining to perturbation theory to a static potential $\mathbf{A}$.
A constant, time-independent vector potential generates no electric
or magnetic field, therefore adding a constant time-independent gauge
potential is no different than changing gauge. For systems displaying
gauge invariance, this term is therefore expected to vanish. In the
thermodynamic limit where this condition is satisfied, we can write
\[
\begin{aligned}\text{Tr}\left(\rho^{\left(0\right)}\Delta^{ij}\right) & =-\sum_{\left\{ \varepsilon_{\lambda}=\varepsilon_{\lambda'}\right\} }\frac{df_{\lambda}}{d\varepsilon}v_{j}^{\lambda\lambda'}v_{i}^{\lambda'\lambda}\\
 & -\sum_{\left\{ \varepsilon_{\lambda}\neq\varepsilon_{\lambda'}\right\} }\left(\frac{f_{\lambda}-f_{\lambda'}}{\varepsilon_{\lambda}-\varepsilon_{\lambda'}}\right)v_{j}^{\lambda\lambda'}v_{i}^{\lambda'\lambda}.
\end{aligned}
\]
Since $A\left(\omega\right)=E\left(\omega\right)/\left(\omega+i0^{+}\right)$,
the full expression for the first order correction to the current
is therefore
\[
\begin{aligned}\left\langle \mathcal{J}_{i}\right\rangle \left(\omega\right)=-iE_{j}\left(\omega\right)\int_{FBZ}d\mathbf{k}\left\{ \sum_{\left\{ \varepsilon_{\lambda}=\varepsilon_{\lambda'}\right\} }\frac{1}{\omega+i\gamma}\frac{df_{\lambda}}{d\varepsilon}v_{j}^{\lambda\lambda'}v_{i}^{\lambda'\lambda}\right.\\
\left.+\frac{e}{\hbar}\sum_{\left\{ \varepsilon_{\lambda}\neq\varepsilon_{\lambda'}\right\} }\frac{f_{\lambda}-f_{\lambda'}}{\omega_{\lambda\lambda'}\left(\omega+\omega_{\lambda\lambda'}+i\gamma\right)}v_{j}^{\lambda\lambda'}v_{i}^{\lambda'\lambda}\right\} .
\end{aligned}
\]
The first term is coined the Drude term and the second describes interband
dynamics.

\subsection{Matsubara Calculation for $\rho^{\text{eq}}$\label{subsec:Matsubara-Calculation-for}}

\[
\rho^{\text{eq}}\left(t\right)=f\left(H_{0}+V\left(t\right)\right)
\]
To calculate this quantity to first order in $\mathbf{A}\left(t\right)$,
we may use the Matsubara sum. To do so, we consider that at each time
$t$ we have a time-independent Hamiltonian. The Fermi-Dirac distribution
can be written as 
\[
f(\varepsilon)=\frac{1}{\beta}\sum_{i\omega_{n}}\frac{1}{i\omega_{n}-\varepsilon}
\]
Where $\omega_{n}=\frac{2\pi}{\beta}\left(n+1/2\right)$ are the fermionic
Matsubara frequencies that correspond to the simple poles of the Fermi-Dirac
distribution $f\left(i\omega_{n}+z\right)=\frac{1}{e^{\beta i\omega_{n}}e^{\beta z}+1}=\frac{1}{-e^{\beta z}+1}\simeq-\frac{1}{\beta z}$.
We may extend it to be a function of the Hamiltonian and approximate
it to first order in $V(t)$: 
\[
\begin{aligned}f(H)= & \frac{1}{\beta}\sum_{i\omega_{n}}G_{0}\left(i\omega_{n}\right)\left\{ 1+V\left(t\right)G_{0}\left(i\omega_{n}\right)\right\} +O\left(V\left(t\right)^{2}\right)\end{aligned}
\]
Where $G_{0}\left(i\omega_{n}\right)\equiv\left(i\omega_{n}-H_{0}\right)^{-1}$
is the Green function of the free system evaluated at $i\omega_{n}$.
Therefore, we have the zeroth and first terms of the Fermi-Dirac distribution
wich can be written in the eigenbasis of $H_{0}$ as 
\begin{equation}
\begin{alignedat}{1}\rho_{\alpha\beta}^{\text{eq}\left(0\right)}\left(t\right) & =\frac{1}{\beta}\sum_{i\omega_{n}}\frac{1}{i\omega_{n}-\varepsilon_{\alpha}}\delta_{\alpha\beta},\\
\rho_{\alpha\beta}^{\text{eq}\left(1\right)}\left(t\right) & =\frac{1}{\beta}V_{\alpha\beta}\left(t\right)\sum_{i\omega_{n}}\frac{1}{i\omega_{n}-\varepsilon_{\alpha}}\frac{1}{i\omega_{n}-\varepsilon_{\beta}}.
\end{alignedat}
\label{eq:mats_terms}
\end{equation}
Because $\omega_{n}$ are the poles of the Fermi-Dirac distribution,
we can write each term in the sum of Eq. \ref{eq:mats_terms} as a
closed path integral around the pole 
\[
2\pi i\frac{1}{i\omega_{n}-\varepsilon_{\alpha}}\frac{1}{i\omega_{n}-\varepsilon_{\beta}}=-\beta\oint_{\gamma_{n}}dzf\left(z\right)\frac{1}{z-\varepsilon_{\alpha}}\frac{1}{z-\varepsilon_{\beta}},
\]
Where $\gamma_{n}$ denotes the path around the pole $i\omega_{n}$.
Notice the integrand has singularities coming from the Fermi-Dirac
distribution and from the term $\frac{1}{z-\varepsilon_{\alpha}}\frac{1}{z-\varepsilon_{\beta}}$.
Since the integrand goes to zero for $\left|z\right|\rightarrow+\infty$,
we may say that the total sum of all residues is zero. This means
that the sum in the poles of the Fermi-Dirac distribution equals the
sum of the residues of the singularities coming from $\frac{1}{z-\varepsilon_{\alpha}}\frac{1}{z-\varepsilon_{\beta}}$.
This allows us to write
\[
\rho_{\alpha\beta}^{\text{eq}\left(1\right)}\left(t\right)=\frac{V_{\alpha\beta}\left(t\right)}{2\pi i}\left(\oint_{\gamma_{\alpha}}+\oint_{\gamma_{\alpha}}\right)dzf\left(z\right)\frac{1}{z-\varepsilon_{\alpha}}\frac{1}{z-\varepsilon_{\beta}},
\]
Which gives
\begin{equation}
\begin{alignedat}{1}\rho_{\alpha\beta}^{\text{eq}\left(1\right)}\left(t\right) & =\begin{cases}
\left(\frac{f\left(\varepsilon_{\alpha}\right)-f\left(\varepsilon_{\beta}\right)}{\varepsilon_{\alpha}-\varepsilon_{\beta}}\right)V_{\alpha\beta}\left(t\right), & \varepsilon_{\alpha}\neq\varepsilon_{\beta},\\
\frac{df\left(\varepsilon_{\alpha}\right)}{d\varepsilon}V_{\alpha\beta}\left(t\right), & \varepsilon_{\alpha}=\varepsilon_{\beta}.
\end{cases}\end{alignedat}
\label{eq:rho_eq}
\end{equation}

\section{Aknowledgements}

This work was supported by Fundação para a Ciência e a Tecnologia
(FCT, Portugal) in the framework of the Strategic Funding UID/04650/2025
- Centro de Física das Universidades do Minho e do Porto. Further
support from Fundação para a Ciência e a Tecnologia (FCT, Portugal)
through project No. EXPL/FISMAC/0953/2021 (B.A. and J.M.A.P.), and
Grants. No. 2023.02155.BD (J.M.A.P), and No. CEECIND/02936/2017 (B.A.)
is acknowledged.

\bibliographystyle{apsrev4-2}
\bibliography{refs_exact_sol}

\end{document}